\documentclass[useAMS,usenatbib]{aa}  

\usepackage{graphicx}
\usepackage{txfonts}
\usepackage{xcolor}

\usepackage{hyperref}
\usepackage{subfigure}

\newcommand{\srr}[1]{{\color{black}#1}} 
\graphicspath{{images/}}

\AtBeginDocument{\mathcode`v=\varv}

\begin{document}

   \title{Cosmic ray ionisation of a post-impact early Earth atmosphere}

   \subtitle{Solar cosmic ray ionisation must be considered in origin-of-life scenarios}

   \author{S. R. Raeside
          \inst{1,2}
          \and
          D. Rodgers-Lee
          \inst{1}
          \and
          P. B. Rimmer
          \inst{3}
          }

    \institute{Astronomy \& Astrophysics Section, School of Cosmic Physics, Dublin Institute for Advanced Studies, 31 Fitzwilliam Place, Dublin D02 XF86, Ireland
    \and{School of Physics, Trinity College Dublin, The University of Dublin, College Green, Dublin 2, Ireland}
    \and{Cavendish Laboratory, University of Cambridge, JJ Thomson Ave, Cambridge, CB3 0HE, United Kingdom}\\
    \email{shaunarose@cp.dias.ie}}

   \date{Received 1 November 2024 / Accepted 2 April 2025}
 
  \abstract
   {Cosmic rays, both \srr{solar} and Galactic, have an ionising effect on the Earth's atmosphere and are thought to be important in the production of prebiotic molecules. In particular, the $\rm{H_2}$-dominated atmosphere that follows an ocean-vaporising impact is considered favourable to prebiotic molecule formation. As a first step in determining the role that cosmic rays might have played in the origin of life we need to understand the significance of their ionising effect.}
   {We model the transport of \srr{solar} and Galactic cosmic rays through a post-impact early Earth atmosphere at 200 Myr. We aim to identify the differences in the resulting ionisation rates - particularly at the Earth's surface during a period when the Sun was very active.}
   {We use a Monte Carlo model for describing cosmic ray transport through the early Earth atmosphere, giving the cosmic ray spectra as a function of atmospheric height. Using these spectra we calculate the ionisation and ion-pair production rates as a function of height due to Galactic and \srr{solar} cosmic  rays. The Galactic and \srr{solar} cosmic  ray spectra are both affected by the Sun's rotation rate, $\Omega$, because the solar wind velocity and magnetic field strength both depend on $\Omega$ and influence cosmic ray transport. We consider a range of input spectra resulting from the range of possible rotation rates of the young Sun - from \srr{$3.5-15\, \Omega_{\rm{\odot}}$}. To account for the possibility that the Galactic cosmic ray spectrum outside the Solar System is not constant over Gyr timescales, we compare the ionisation rate at the top of the Earth's atmosphere resulting from two different scenarios. We also consider the suppression of the cosmic ray spectra by a planetary magnetic field.}
   {We find that the ionisation and ion-pair production rates due to cosmic rays are dominated by \srr{solar} cosmic  rays in the early Earth atmosphere for most cases. The corresponding ionisation rate at the surface of the early Earth ranges from $5 \times 10^{-21}\rm{s^{-1}}$ for $\Omega = 3.5\,\Omega_{\rm{\odot}}$ to \srr{$1 \times 10^{-16}\rm{s^{-1}}$ for $\Omega = 15\,\Omega_{\rm{\odot}}$}. Thus if the young Sun was a fast rotator \srr{($\Omega = 15\,\Omega_{\rm{\odot}}$)}, it is likely that \srr{solar} cosmic rays had a significant effect on the chemistry at the Earth's surface at the time when life is likely to have formed.}
   {Cosmic rays, particularly \srr{solar} cosmic  rays, are a source of ionisation that should be taken into account in chemical modelling of the post-impact early Earth atmosphere. Modelling of cosmic ray transport and effects on chemistry will also be of interest for \srr{the} characterisation of $\rm{H_2}$-dominated exoplanet atmospheres.}

   \keywords{cosmic rays --
                Earth -- Sun: particle emission -- 
                methods: numerical -- Planets and satellites: atmospheres
               }

   \maketitle

\section{Introduction}
\label{sec:introduction}
The origin of life depends on many factors including the presence of a number of key molecules, such as hydrogen cyanide (HCN) and formaldehyde (HCHO), that act as “building blocks” for life and are known as prebiotic molecules \citep{book-gargaud,benner-2020}. Cosmic rays - energetic charged particles - are thought to be important in creating the conditions necessary to form these prebiotic molecules in a planetary atmosphere \citep[e.g.][]{airapetian-2016}. 

\srr{Solar cosmic  rays (i.e. particles accelerated \srr{by flares} and in coronal mass ejections, also known as solar energetic particles)} and Galactic cosmic rays, which will be collectively referred to as `cosmic rays' throughout, contribute to ionisation in the Earth's atmosphere \citep[e.g.][]{sinnhuber-2012}. If the rate of cosmic rays impacting the atmosphere is large enough, this ionisation will significantly alter the chemical species present. 

Molecules such as HCN \citep{benner-2019} are described as prebiotic molecules because they are needed to produce the amino acids and RNA bases that go on to form cells - the basic unit of life \citep{book-gargaud}. The production of many of these molecules appears to require an oxygen-poor environment \citep{benner-2019}. Experiments such as those described in \cite{miller-1983} highlight the need for an oxygen-poor atmosphere with large amounts of molecular hydrogen ($\mathrm{H_2}$). \cite{zahnle-2020} describe how a transient oxygen-poor, $\mathrm{H_2}$-rich atmosphere could have been produced on Earth as a result of \srr{a large} impact early in Earth's lifetime. \srr{\citet{zahnle-2020} present this post-impact early Earth atmosphere scenario} as an explanation for the observed excess of metals in the Earth's mantle which are typically dissolved in iron and would otherwise be expected to be in the core. A single impactor large enough to deliver all of the excess of these metals would be comparable in size ($\sim 2300$ km diameter) to Pluto \citep{zahnle-2020}. An impact of this size would likely melt and ‘reset’ the crystals used for radiogenic dating of rock in Earth’s crust \citep{benner-2020}. The existence of material in the Earth’s crust which has been dated to $>$ 4.35 Gyr ago implies that the transient post-impact atmosphere would have been present when Earth was only a few hundreds of millions of years old \citep{brasser-2016,benner-2020}.

The properties of the young Sun produce differences in the intensity and energy of cosmic rays reaching Earth compared with the present day. From photometric observations of Sun-like stars \citep{gallet-bouvier-2013}, which show faster rotation rates at younger ages, we can assume that the Sun's rotation rate was faster at earlier epochs than at present. The observed scaling of the large-scale magnetic field strength with the rotation rate of low-mass stars \citep{vidotto-2014} indicates that the young Sun also had a stronger magnetic field than today. This stronger magnetic field \srr{and the increased flaring rates of fast rotating low-mass stars \citep{gunther-2020}} lead to increased acceleration of \srr{solar} cosmic rays \citep{drl-2021} and likely increased the suppression of Galactic cosmic rays by the \srr{solar wind} \srr{at earlier ages} \citep{drl-2020}. 

Some of the first origin of life studies investigated possible energy sources leading to the synthesis of prebiotic molecules. \cite{miller-urey-1959} designed experiments to simulate early Earth atmospheres and investigate the efficacy of lightning and solar UV radiation as energy sources for amino acid production. \cite{miller-urey-1959} suggested that Galactic cosmic rays were unlikely to have a large effect on the atmosphere’s chemistry because the cosmic ray energy density was likely to be negligible in the past. However, they did not consider \srr{solar} cosmic  rays, which are likely to have been present at Earth in greater numbers in the past. More recent experiments \citep{kobayashi-2023} revisited the question of cosmic rays’ role in forming prebiotic molecules in an early Earth atmosphere, taking into account both Galactic and \srr{solar} cosmic  rays. The results of these experiments point toward \srr{solar} cosmic  rays as the most promising energy sources for prebiotic molecule production.

\srr{More broadly,} since it is not possible to observe the early Earth atmosphere, spectroscopic observations of exoplanets with different atmospheres in a variety of cosmic ray environments can provide insight into the conditions which promote the production of prebiotic molecules \citep{rimmer-2023}. Cosmic rays can also negatively impact on the likelihood of life on exoplanets - \citet{herbst-2024} modelled the effect of cosmic rays on Earth-like exoplanet atmospheres, finding that cosmic rays can destroy biosignature molecules, such as ozone, which suggest the presence of life. \citet{drl-2023} modelled cosmic ray transport in the atmosphere of GJ436b, an exoplanet with an atmosphere comparable to the early Earth which was observed with NIRCam in JWST Cycle 1. Transmission spectroscopy observations of such exoplanets will improve our understanding of origin-of-life scenarios. 

 \citet{rimmer-2013} found that cosmic ray ionisation is important to consider in chemical modelling of brown dwarf and free-floating exoplanet atmospheres. \citet{rimmer-2016} modelled \srr{cosmic ray transport} through exoplanet atmospheres including a hydrogen-dominated gas giant. Chemical models such as that presented in \citet{rimmer-2016} use the cosmic ray ionisation rate as an important input for modelling atmospheric chemistry, including \srr{prebiotic molecule production}. As a starting point, \citet{rimmer-2016} used the same top-of-atmosphere cosmic ray spectra for each scenario. \srr{Here,} our cosmic ray spectra \srr{depend} on the solar rotation rate and reflect the effects of transport through the \srr{solar wind} on the spectra. \citet{drl-2021} investigated the cosmic ray \srr{intensity} at the top of the Earth’s atmosphere at an age of 600 Myr. 
 
 Here we investigate the \srr{transport and ionising effect of cosmic rays} through the Earth’s atmosphere at 200 Myr, when the oxygen-poor atmosphere required for prebiotic chemistry is thought to have been present. We consider two different scenarios for the Galactic cosmic ray spectrum outside the Solar System which may not be constant on Gyr timescales. We also modify our cosmic ray spectra for the post-impact early Earth atmosphere to account for deflection by a planetary magnetic field. The paper is structured as follows: Section\,\ref{sec:methodology} introduces the models that are used for the cosmic ray transport and in Section\,\ref{sec:results} our results are presented. The discussion and conclusions are given in Sections\,\ref{sec:discussion} and \ref{sec:conclusions}.
\section{Methodology}
\label{sec:methodology}
Here we describe how we have modelled the cosmic ray transport through the early Earth's atmosphere. Section \ref{subsec:density-profile} presents the model properties of the atmosphere. \srr{Section \ref{subsec:stellar-wind} presents the solar wind properties for the early Earth scenario. Section \ref{subsec:solar-transport}} details the role of the solar wind in determining the cosmic ray spectra reaching the top of the atmosphere, which are then presented in Section \ref{subsec:fluxes}. Section \ref{subsec:cr-model} introduces the model used for cosmic ray transport.
\subsection{Post-impact early Earth atmosphere model}
\label{subsec:density-profile}
In this paper we consider the early Earth\srr{'s} atmosphere at 200 Myr, after the impact of a Pluto-sized dwarf planet rich in iron. In this scenario the impactor vaporises the Earth's oceans, based on the "maximum late veneer" scenario described in \citet{zahnle-2020}. The water from the vaporised oceans reacts with the iron from the impactor, producing $\rm{H_2}$ and $\rm{FeO}$ through the reaction $\rm{ Fe+H_2O\rightarrow H_2 + FeO }$. The resulting atmosphere is hydrogen-dominated and has a significantly higher pressure at the surface than the present-day, ranging from several tens of bars \citep{zahnle-2020} to $\sim 120$ bar \citep{itcovitz-2022}. We will use the terms "post-impact early Earth" and "early Earth" interchangeably to refer to this scenario, except where stated otherwise. 

\begin{figure}
   \centering
   \includegraphics[width=\hsize]{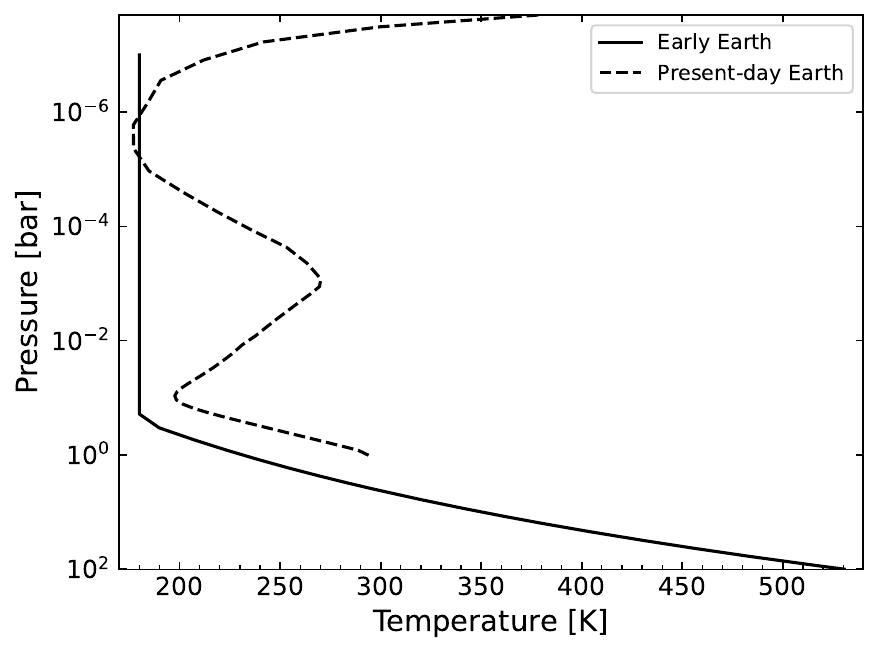}
      \caption{Temperature-pressure profile for a post-impact early Earth atmosphere\srr{, with the present-day temperature-pressure profile for comparison}. See Section \ref{subsec:density-profile} for details.
              }
         \label{fig:tp-profile}
   \end{figure}

The temperature-pressure profile we use for the post-impact early Earth atmosphere is shown in Fig. \,\ref{fig:tp-profile}. This temperature-pressure profile was produced by extrapolating the model temperature-pressure profile of Earth's atmosphere at $\sim 800$ Myr \citep[Fig. 1 from][]{tian-2011} to a pressure, $P = 100$ bar. The temperature at the surface is higher than at the present day, exceeding 500 K. The scale height of the atmosphere, $H$, is calculated as $H = k_{\rm{B}}T_{\rm{atm}} / g \mu m_{\rm{p}}$, where $T_{\rm{atm}}$ is the atmosphere temperature, $g$ is acceleration due to gravity, $\mu$ is the atmosphere's mean molecular weight and $m_{\rm{p}}$ is the mass of a proton. \srr{For simplicity in our cosmic ray transport simulations we assume that the atmosphere is 100\% $\rm{H_2}$.} Using $\mu = 2$ and assuming a representative  $T_{\rm{atm}} = 530 \rm{K}$ and constant $g$, \srr{then $H = 210 \rm{km}$ for the post-impact early Earth atmosphere}. For comparison, for Earth's present-day atmosphere $H = 8 \rm{km}$. \srr{ In Fig. \,\ref{fig:tp-profile}, the dashed line represents the temperature-pressure profile for present-day Earth (A. M. Taylor, priv. comm.). For the present-day Earth, the temperature increases with increasing altitude for $P < 10^{-6}$ bar due to absorption of short-wavelength radiation by $\rm{N_2}$ and $\rm{O_2}$, and for $10^{-3}<P < 10^{-1}$ bar due to absorption of solar UV radiation by $\rm{O_3}$. This level of complexity is not included in our model at present.}

\begin{figure}
   \centering
   \includegraphics[width=\hsize]{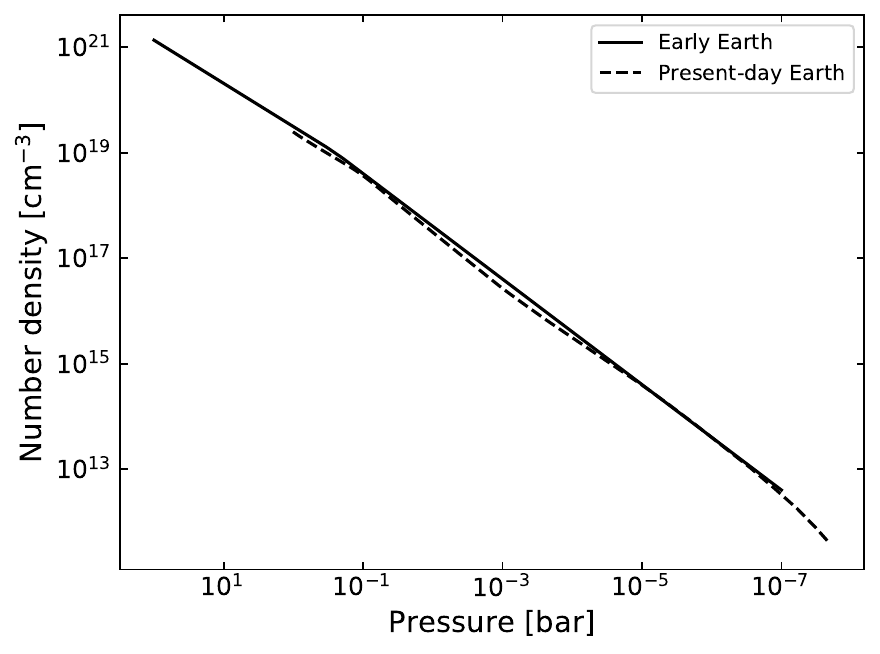}
      \caption{Number density of the atmosphere at 200 Myr (solid line) as a function of pressure, calculated using the ideal gas law and adopted temperature-pressure profile for comparison. The number density for the present-day Earth is shown with the dashed line.
              }
         \label{fig:density}
   \end{figure}

Using the ideal gas law, we calculate the number density, $n$, throughout the atmosphere. The solid line in Fig.\,\ref{fig:density} shows $n$ as a function of $P$ for the post-impact early Earth atmosphere. The present-day density profile is shown for comparison by the dashed line, showing the difference in number density at the surface (1 bar for the present day Earth, 100 bar for the early Earth) of $\sim$2 orders of magnitude. By assuming hydrostatic equilibrium ($dP / dz = - g \mu m_{\rm{p}} n$) we calculate the height, $z$, corresponding to each $P$. \srr{The pressure-altitude profile for the early Earth is shown in Fig\,\ref{fig:pz} in Appendix\,\ref{subsec:pz-plots}.} The density profile uses a logarithmically spaced $z$ grid with 250 points. The early Earth atmosphere density profile is required for the cosmic ray transport model, discussed further in Section \ref{subsec:cr-model}.
\subsection{Solar wind properties}
\label{subsec:stellar-wind}
\begin{table}
\caption{The three different rotation rates of the Sun at 200\,Myr are presented with the associated solar properties used for the solar wind model and cosmic ray transport simulations.}             
\label{table:rotation-rates}      
\centering          
\begin{tabular}{c c c c c c}   
\hline       
 & $\Omega$ & $P_{\rm{rot}}$ & $B_\star$ & $T_\star$ & $n_\star$\\ 
\hline
 Rotation & $[\Omega_{\odot}]$ & [days] & [G] & [MK] & [$\rm{cm^{-3}}$]\\
\hline                    
   Slow & 3.5 & 8 & 7 & 3.2 & $\rm{2 \times 10^8}$\\  
   Medium & 6 & 5 & 14    & 3.8 & $\rm{3 \times 10^8}$\\
   Fast & \srr{15} & \srr{2} & \srr{46}     & \srr{5.4} & \srr{$\rm{5 \times 10^8}$}\\
\hline       
\end{tabular}

\textbf{Notes.} The Sun's rotation rate, $\Omega$, is given in terms of the Sun's present rotation rate, $\Omega_{\odot} = 2.67 \times 10^{-6}\rm{rad \, s^{-1}}$, and the rotation period, $P_{\rm{rot}}$, is given in days. $B_{\star}$ is the radial magnetic field strength at the base of the solar wind, $T_\star$ is the base temperature and $n_\star$ is the base density.
\end{table}

The motion of cosmic rays in the \srr{Solar System} is affected by cosmic ray interactions with the magnetised solar wind. The solar wind properties which influence cosmic ray transport in the \srr{Solar System} \srr{(e.g. the diffusion coefficient calculated using the solar wind magnetic field strength, $B(r)$, and radial velocity, $v_{\rm{r}}(r) \sim v(r)$)} vary with $\Omega$. \srr{For the Sun at 200 Myr, $\Omega$ is unknown. At an age of $\sim 200$ Myr, for low-mass stars (with $M = 0.25 - 1.1 M_{\odot}$) in young clusters the observed rotation rates from photometric observations of stellar surface spots range from $1-100\Omega_{\rm{\odot}}$  \citep{gallet-bouvier-2013}.}

\srr{\citet{johnstone-2021} fit a rotational evolution model to stellar rotation period measurements for $M = 0.9 - 1.1 M_{\odot}$. \citet{johnstone-2021} present rotational evolution tracks for slow, medium and fast rotator cases fit to the 5$^\mathrm{th}$ ($\Omega = 3.5\Omega_{\rm{\odot}}$), 50$^\mathrm{th}$ ($\Omega = 6\Omega_{\rm{\odot}}$) and 95$^\mathrm{th}$ ($\Omega = 15\Omega_{\rm{\odot}}$) percentile of the $\Omega$ measurements, respectively. We use these three different possible rotation rates for the Sun at 200 Myr and model the subsequent effect on cosmic ray transport through the solar wind. However, it is important to note that modelling of losses of sodium and potassium from the Moon's soil due to solar activity \citep{saxena-2019}, comparing to measured abundances in the Moon's soil, suggest that the Sun has evolved as a slow or medium rotator, rather than a fast rotator.}

\srr{Using the magneto-rotator stellar wind model presented in \citet{johnstone-2015} and \citet{carolan-2019}, based on the Versatile Advection Code \citep[VAC][]{
toth_1996}, we obtain $B(r)$ and $v(r)$ along with the solar wind mass loss rate, $\dot{M}$, for a given $\Omega$. Similar to \citet{drl-2020}, we calculate the magnetic field strength ($B_{\rm{\star}}$), temperature ($T_{\rm{\star}}$) and density ($n_{\rm{\star}}$) at the base of the solar wind as a function of $\Omega$ at 200 Myr which are the necessary inputs for the stellar wind model.}

For $B_{\rm{\star}}$, we use the large-scale magnetic field strength of the Sun. This large-scale magnetic field strength is related to $\Omega$ by the \srr{empirical} relation for low-mass stars presented in \citet{vidotto-2014}:\srr{
\begin{equation}
    B_{\star}(\Omega) = 1.3 \left( \frac{\Omega}{\Omega_{\odot}} \right)^{1.32 \pm 0.14} \rm{G}\,.
\end{equation}
\noindent }We use the relationship \srr{for $T_{\rm{\star}}$ as a function of $\Omega$} from \citet{dualta-2018}: \srr{
\begin{equation}
    T_{\star} = \begin{cases} 1.50\,(\frac{\Omega}{\Omega_{\odot}})^{1.2} \,\rm{MK} & \text{for } \Omega < 1.4 \Omega_{\odot}\\ 1.98\,(\frac{\Omega}{\Omega_{\odot}})^{0.37}\, \rm{MK} & \text{for } \Omega \geq 1.4\Omega_{\odot}\end{cases}
\end{equation}

\noindent and calculate $n_{\rm{\star}}$ using $n_{\rm{\star}} = 10^8 (\Omega / \Omega_{\rm{\odot}})^{0.6}\rm{cm^{-3}}$ from \citet{ivanova-2003}.} For 200 Myr, $B_{\rm{\star}}$ ranges from \srr{$7 - 46$} G and $v$ at 1~au ranges from \srr{$820 - 1200 \rm{kms^{-1}}$}. \srr{In comparison,} at the present day $B_{\rm{\star}} = 1.3$ G\srr{, in agreement with the observed magnetic field strength of the dipolar component of the Sun averaged over solar cycles 21 to 23 \citep[see Fig. 1 in ][]{johnstone-2015},} and $v$ at 1au ranges from $400 - 600 \rm{kms^{-1}}$ \srr{\citep{mccomas-2008}}. \srr{These values are also given in Table\,\ref{table:rotation-rates}, along with the rotation period.} 

The radius of the heliosphere \srr{(i.e. the heliopause distance)}, $R_{\rm{h}}$, is the outer boundary for our modelling of cosmic ray propagation through the solar wind. The heliosphere is the cavity in the interstellar medium (ISM) carved out by the solar wind. Table \ref{table:results} presents $v_{\rm{1 au}}$, $\dot{M}$ and $R_{\rm{h}}$ at 200 Myr for the slow, medium and fast solar rotation rates. $R_{\rm{h}}$ is determined by the radial distance from the Sun where the solar wind ram pressure \srr{($\dot{M}v / r$)} is balanced with the ambient ISM pressure ($P_{\rm{ISM}}$). Following \citet{svensmark-2006} in assuming \srr{spherical symmetry and} that $P_{\rm{ISM}}$ is constant as a function of time, we have calculated $R_{\rm{h}}$ as
\begin{equation}
    R_{\rm{h}}(\Omega) = R_{\rm{h,\odot}}\sqrt{\frac{\dot{M}(\Omega)v(\Omega)}{\dot{M}_{\odot}v_{\odot}}} \, ,
    \label{eq:Rh}
\end{equation}
where $R_{\rm{h,\odot}}$ is the present-day Sun's heliospheric radius, taken to be 122 au \citep{vos-2015}. We find that at 200 Myr \srr{$R_{\rm{h}}=1650-4700$} au for \srr{$\Omega = 3.5 - 15 \Omega_{\rm{\odot}}$}. \srr{ It is possible to use 3D stellar wind modelling of the astrospheres of other stars \citep[see e.g.][]{herbst-2020,engelbrecht-2024,scherer-2024} to perform more detailed calculations of the astrospheric distance, if the ISM properties are known. However, given that little is known about the ISM properties during the post-impact early Earth scenario Eq.\,\ref{eq:Rh} currently appears sufficient.} The inner boundary for our model is \srr{$r = 0.005$au ($\sim R_\odot$)}. 

\subsection{Cosmic ray transport in the \srr{Solar System}}
\label{subsec:solar-transport}
To obtain cosmic ray spectra at the top of the early Earth atmosphere we model the transport of solar and Galactic cosmic rays through the solar wind at 200 Myr. \srr{ For this, as a first approach,} we solve the 1D Parker diffusion-advection transport equation \citep{parker-1965} as presented in \citet{drl-2021}:
\begin{equation}
\frac{\partial f}{\partial t} = \nabla \cdot (\kappa \nabla f)-v\cdot \nabla f +\frac{1}{3}(\nabla \cdot v)\frac{\partial f}{\partial \mathrm{ln}p} + Q_\mathrm{inj},
\label{eq:parker}
\end{equation}
where $f(r,p,t)$ is the cosmic ray phase space density, $\kappa(r,p,\Omega)$ is the spatial diffusion coefficient, $v(r,\Omega)$ is the radial velocity of the stellar wind and $p$ is the momentum of the cosmic rays \srr{with $p = 0.1 - 300\, \mathrm{GeV} / c$}. The injection of solar cosmic rays accelerated by solar flares is represented by $Q_\mathrm{inj}$ and varies as a function of $\Omega$ \srr{which is discussed further in Section\,\ref{subsec:fluxes}. The cosmic ray differential intensity (i.e. the number of cosmic rays per unit area, steradian, time and kinetic energy) is related to the phase space density by $j(T) = p^2 f(p)$, where $T$ is the cosmic ray kinetic energy. }

For a given level of turbulence in the solar wind the solar wind magnetic field strength, $B$, determines the diffusion coefficient of cosmic rays: for larger values of $\kappa$ cosmic rays travel further before being scattered. Similar to \citet{drl-2020}, we assume a level of turbulence which is constant with $\Omega$. This means that $\kappa$ can be expressed as $\kappa / \beta c = r_{\rm{L}}$, where $\beta$ is the ratio of particle speed relative to the speed of light and $r_{\rm{L}} = p/eB$ is the \srr{ cosmic ray's} Larmor radius. The diffusion coefficient decreases with increasing $B$ and $\Omega$. 

The momentum advection of cosmic rays depends on the divergence of the solar wind as cosmic rays lose energy through doing work against the expanding solar wind \citep{parker-1965}. Spatial advection is determined by the solar wind velocity and affects Galactic and solar cosmic rays differently due to the source location: for Galactic cosmic rays advection out of the \srr{Solar System} suppresses the \srr{cosmic ray fluxes} entering the \srr{Solar System}, while solar cosmic rays are simply advected out from the Sun through the \srr{Solar System} by the solar wind.  

\srr{ It is important to note that 3D stellar wind and Galactic cosmic ray transport models have been applied to a number of exoplanetary systems \citep{engelbrecht-2024,scherer-2024,light-2025} and include 3D Galactic cosmic ray transport effects that are not captured by 1D models, such as anisotropic diffusion and 3D particle drifts. However, applying the same methodology to the young Sun remains challenging. For instance, it is difficult to construct a 3D stellar wind model given that we do not have a magnetic map  for the young Sun \citep[as are available for other low-mass stars, e.g.][]{bellotti-2023} which is an important input for any 3D solar/stellar wind models.} The solar wind properties presented in Table \ref{table:results} are used to model the cosmic ray transport through the \srr{Solar System} and produce a range of top-of-atmosphere spectra at 1 au at 200 Myr \srr{that} are presented \srr{next}. 
   \begin{figure*}
   \centering
\subfigure[]{
   \includegraphics[width=0.45\textwidth]{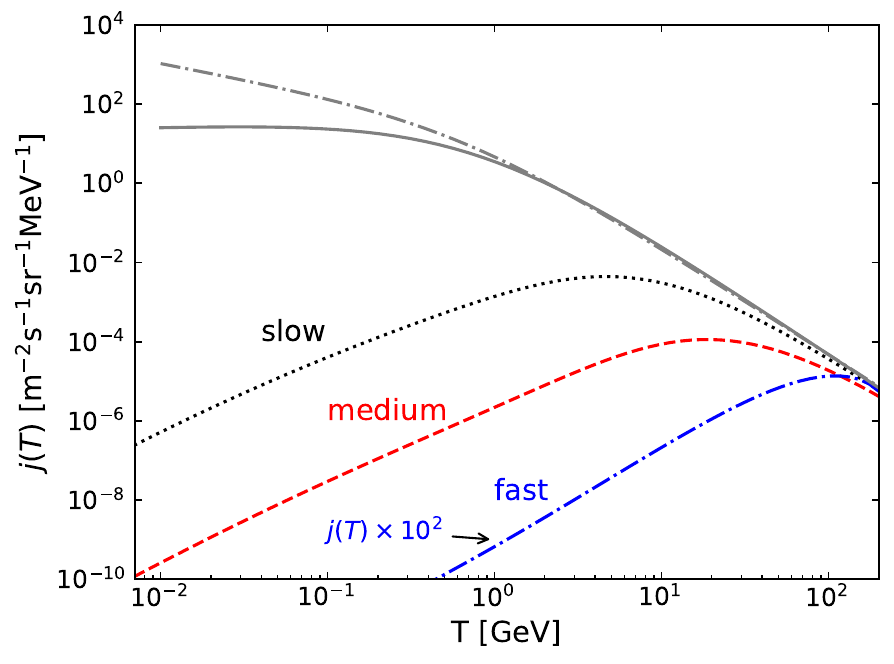}
\label{fig:TOA_gcr}}  
\centering
\subfigure[]{
   \includegraphics[width=0.45\textwidth]{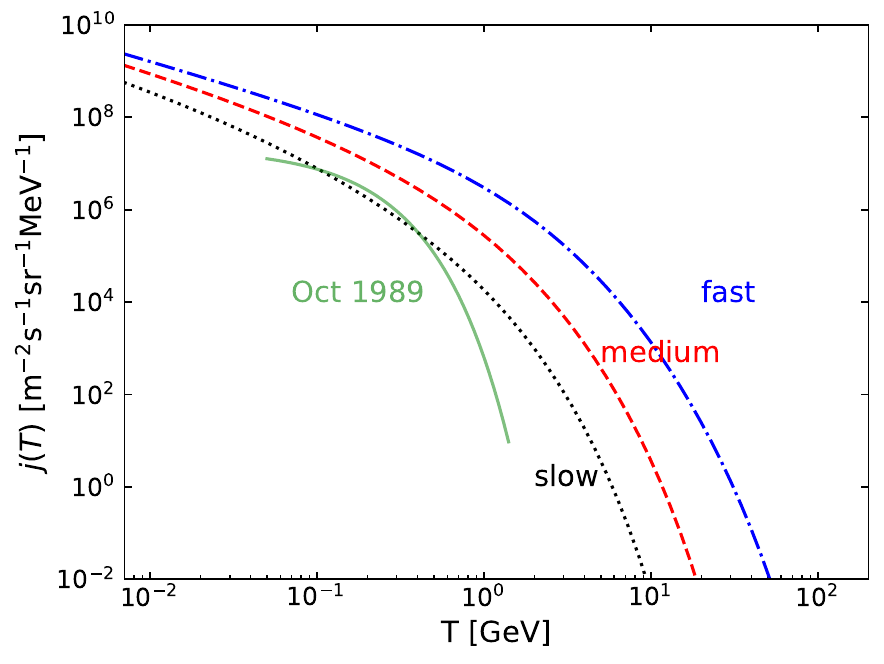}
\label{fig:TOA_scr}}  
   \caption{(a) Top-of-atmosphere Galactic cosmic ray spectrum at 1au at 200Myr. Spectra for a slow (black dotted line), medium (red dashed line) and fast (blue dot-dashed line) solar rotation rate. The LIS is shown by the solid grey line. The enhanced LIS shown by the dot-dashed grey line is discussed in Section \ref{subsec:vary-lis}. The spectrum for the fast rotation rate has been multiplied by \srr{$10^2$} to allow all of the spectra to be shown within a reasonable range on the same axes. (b) Top-of-atmosphere solar cosmic ray spectrum at 1au at 200Myr for a slow, medium and fast solar rotation rate. The linestyles represent the same scenarios as in the Galactic cosmic ray spectra. \srr{The green line represents the fit of the observed spectrum \citet{reeves-1992} of the solar energetic particle event measured at Earth in October 1989.}}
              \label{fig:TOA}
    \end{figure*}
\subsection{Top-of-atmosphere cosmic ray spectra}
\label{subsec:fluxes}

Modelling the cosmic ray transport within the atmosphere requires initial spectra at the top of the atmosphere. As a starting point we do not account for the effect of a planetary magnetic field, which would deflect lower-energy cosmic rays toward the poles. This effect is discussed further in Appendix \ref{subsec:bfield}.

For Galactic cosmic rays the top-of-atmosphere spectra are acquired by modelling \srr{cosmic ray transport} through the \srr{Solar System}, starting with the Galactic cosmic ray spectrum outside of the heliosphere. The local interstellar spectrum (LIS) refers to the \srr{Galactic cosmic ray spectrum} containing contributions from sources within thousands of au from the Sun which may be excluded from average Galactic spectra \citep{potgieter-2013}. \srr{There are numerous LIS models based on \textit{Voyager 1} and \textit{2} data at different distances as they moved through and out of the heliosphere \citep[e.g.][]{vos-2015,corti-2016,herbst-2017,engelbrecht-2021}. } The LIS we use is derived from \textit{Voyager 1} data taken at the edge of the heliosphere, given by Eq.\,1 of \citet{vos-2015}:
\begin{equation}
\label{eqn:vos}
    j_{\rm{LIS}} = 2.70 \frac{T^{1.12}}{\beta^2} \left( \frac{T + 0.67}{1.67} \right)^{-3.93}\rm{MeV^{-1}s^{-1}m^{-2}sr^{-1}} \, .
\end{equation}

The solid grey line in Fig.\,\ref{fig:TOA_gcr} shows the LIS, which we assume to be constant over Gyr time scales. An enhanced LIS is shown by the grey dot-dashed line and will be discussed in Section \ref{subsec:vary-lis}. The black dotted, red dashed and blue dot-dashed lines represent the calculated spectra at Earth at 200 Myr for the slow, medium and fast rotator cases, respectively. The spectrum for the fast rotation rate has been multiplied by \srr{$10^2$} to allow all of the spectra to be shown within a reasonable range on the same axes. The Galactic cosmic ray flux is lower at Earth at 200 Myr than in the local ISM, with a larger difference at lower energies. The reduction in $j$ between the ISM and Earth is greater for higher solar rotation rates. For all values of $\Omega$ considered here, $j$ at 1au is also significantly lower than the present-day values \citep[e.g. Fig. 3 from][]{vos-2015}, particularly at low energies.

\srr{\citet{engelbrecht-2024} present 3D Galactic cosmic ray transport modelling for the astrosphere of Prox Cen, and find that $\Omega$ additionally influences the 3D transport effects not captured in 1D modelling. For Prox Cen \citet{engelbrecht-2024} find that the underwound stellar wind magnetic field leads to the Galactic cosmic ray transport dominated by diffusion parallel to the magnetic field. The diffusion coefficient is larger for parallel than for perpendicular diffusion, which leads to increased Galactic cosmic ray flux in this slow rotation scenario. For the early Earth with a faster rotation rate than the present day we would expect the solar wind's magnetic field to be more tightly wound and for inward Galactic cosmic ray transport to be hindered even more than is shown in Fig.\,\ref{fig:TOA_gcr} by the increased contribution of diffusion perpendicular to the magnetic field.}

The solar cosmic ray spectra at the top of Earth's atmosphere at 200 Myr are shown in Fig. \ref{fig:TOA_scr} for the slow (black dotted line), medium (red dashed line) and fast (blue dot-dashed line) rotator cases. In the case of solar cosmic rays, $j$ is higher over the entire energy range for a faster solar rotation rate, in contrast to Galactic cosmic rays (Fig. \ref{fig:TOA_gcr}, note the different $y$ axis scales). 

The changes in the solar cosmic ray spectrum with \srr{$\Omega$ are mainly due to the initial spectra close to the solar surface being different. Using the same approach as \citet{drl-2021}, the initial solar cosmic ray spectra depend on the solar wind properties and $B_\star$. The spectrum is such that $\frac{d\dot{N}}{dp} \propto p^{-\alpha}\,e^{-p/p_\mathrm{max}}$ where $\dot{N}$ is the number of particles injected per unit time and $p_\mathrm{max}$ is the maximum momentum that the Sun accelerates particles to. We use $\alpha = 2$, compatible with diffusive shock acceleration \citep[e.g.][]{bell-1978} or acceleration due to magnetic reconnection. We determine $p_\mathrm{max}$ using $B_\star$ and the Hillas criterion \citep{hillas-1984}, given by Eq.\,7 in \citet{drl-2021}. We find $p_\mathrm{max}$ ranges from $p_\mathrm{max} = 1.1 \mathrm{GeV} / c$ in the slow rotator case to \srr{$p_\mathrm{max} = 7.1$ GeV$/ c$} in the fast rotator case. Finally, to normalise the spectrum, we assume that $L_{\rm{CR}} \sim 0.1 P_{\rm{SW}}$ where $P_{\rm{SW}} = \dot{M}(\Omega)v_{\rm{1au}}(\Omega)^2/2$ is the kinetic power in the solar wind and $L_{\rm{CR}}$ is the total injected kinetic power in solar cosmic rays such that
\begin{equation}
L_\mathrm{CR} = \int\limits^{300\mathrm{GeV}/c}_{0.1 \mathrm{GeV}/c} \frac{d\dot{N}}{dp}\,T(p)\,dp.
\end{equation}
\noindent Overall, this means that as $\Omega$ increases, the corresponding increase in $B_\star$ and $P_\mathrm{SW}$ lead to higher solar cosmic ray fluxes and higher maximum solar cosmic ray energies. 

In Fig.\,\ref{fig:TOA_scr} we include a comparison to a significant solar event with measured solar energetic particle fluxes (green line). It represents the fit by \citet{reeves-1992} to observations of proton fluxes at geosynchronous orbit during a significant solar energetic particle event in October 1989 \citep[Day 293 in Fig. 5 of][]{reeves-1992}. For $2\times10^{-1}<\textit{T}<5 \times 10^{-1} \, \rm{GeV}$, $j$ for the October 1989 event is slightly greater than for our top-of-atmosphere spectrum in the slow rotator scenario. However, at $T = 1\, \rm{GeV}$, $j$ for the October 1989 event is over 1 order of magnitude less than for our slow rotator scenario and approximately 4 orders of magnitude less than for our fast rotator scenario.}

\subsection{Cosmic ray transport in the atmosphere}
\label{subsec:cr-model}
To model the transport of cosmic rays through the early Earth atmosphere we use the Monte Carlo model presented in \citet{rimmer-2012}. With the top-of-atmosphere input cosmic ray spectra (Fig. \ref{fig:TOA}) and atmospheric density profile (Fig. \ref{fig:density}), the cosmic ray transport model distributes a number of cosmic rays, $N$, according to the input cosmic ray spectrum before calculating the energy losses of each cosmic ray over each increment in height above the surface, $dz$, down through the atmosphere. This model is based on the continuous slowing down approximation \citep{padovani-2009} which describes the loss of energy by cosmic rays passing through a column density, $n(z)dz$, of a medium - in this case, the atmosphere of the early Earth. 
The energy lost by each cosmic ray over $dz$ depends on the cosmic ray "optical depth", $\tau = \sigma(T)n dz$, over $dz$, where $\sigma(T)$ is the ionisation cross section.
\srr{For this hydrogen-dominated atmosphere the $\rm{H_2}$ ionisation cross section described in Eqs. 5 - 6 of \citet{padovani-2009}, from the fitting of experimental data presented in \citet{rudd-1985}, is the dominant $\sigma$}. \srr{The ionisation cross section of $\rm{H_2}$ is the only cross section we consider in our simulation setup. }For the cosmic ray proton energy range considered here, $6.9\times10^{-3} \leq \textit{T} \leq262\,\rm{GeV}$, $\sigma$ is larger for lower-energy cosmic rays \citep[see ][Fig. 1]{padovani-2009}. 

We use $W$ to refer to the average \srr{cosmic ray} energy loss in an ionising collision with $\rm{H_2}$ taken from Eq. 3 of \citet{rimmer-2012}:
\begin{equation}
    W = 7.92 T^{0.82} + 4.76, \,
\end{equation}
with $T$ in eV. The energy loss of each cosmic ray proton is calculated by assigning a random number, $A$, between 0 and 1 to each cosmic ray and comparing $\tau$ with both the energy of the proton and $A$ as follows:
\begin{enumerate}
    \item If $A <\tau<1$, the cosmic ray loses energy $W$.
    \item If $1\leq \tau < \frac{T}{W}$, the cosmic ray loses a random amount of kinetic energy between 0 and $\tau W$.
    \item If $\tau\geq\frac{T}{W}$, the cosmic ray loses all of its kinetic energy and does not contribute to the cosmic ray flux at the next $dz$.
\end{enumerate}
In our modelling of cosmic rays in the early Earth atmosphere we use $N = 500000$ and $10^{-3}\rm{km}\leq \textit{z} \leq 2700 \rm{km}$, with a logarithmically spaced $z$ grid.
The result of these calculations is a set of values of $j$ at each energy value for each of the 250 heights in the atmosphere.\\
We use $j(T,z)$ to calculate the important quantities for chemical modelling of the early Earth atmosphere - the ionisation rate of $\rm{H_2}$ and the ion-pair production rate, $Q(z)$, discussed in Section \ref{subsec:ionisation}. We calculate the ionisation rate of $\rm{H_2}$, $\zeta$, using:
\begin{equation}
      \zeta(z) = 4\pi\int_{I(\rm{H_2})}^{T_{\rm{max}}}j(T,z)[1 + \phi(T)]\sigma(T)dT, \,
      \label{eq:zeta}
   \end{equation}
where $I(\rm{H_2})$ is the ionisation potential of $\rm{H_2}$ and $\phi(T)$ is an energy-dependent correction factor to account for ionisation by secondary electrons \citep[e.g.][]{drl-2023}.
Our results for $j(T,z)$ and $\zeta(z)$ as a function of height in the early Earth atmosphere are presented in Section \ref{sec:results}.
\begin{table}
\caption{The solar wind model output properties are presented with the cosmic ray ionisation rates for the post-impact early Earth atmosphere at 200 Myr.  }             
\label{table:results}      
\centering    
\begin{tabular}{c c c c c c}   
\hline       
 & $v_{\rm{1 au}}$ & $\dot{M}$ & $R_{\rm{h}}$ & $\rm{\zeta^{\rm{TOA}}}$ & $\rm{\zeta^{surf}}$\\ 
\hline
 Rotation & $\rm{[km\, s^{-1}]}$ & $[\dot{M_{\odot}}]$ & [au] & $\rm{[s^{-1}]}$ & $\rm{[s^{-1}]}$\\
\hline                    
   Slow & 820 & 99 & 1640 & $\rm{3 \times 10^{-11}}$ & $\rm{5 \times 10^{-21}}$\\  
   Medium & 940 & 210 & 2580 & $\rm{8 \times 10^{-11}}$ & $\rm{8 \times 10^{-20}}$\\
   Fast & \srr{1200} & \srr{560} & \srr{4700} & $\rm{2 \times 10^{-10}}$ & \srr{$\rm{1 \times 10^{-16}}$} \\
\hline                  
\end{tabular}

\textbf{Notes.} For each of the three assumed solar rotation rates the solar wind velocity at 1 au, \srr{$v_{\rm{1 au}}$}, and mass loss rate, $\dot{M}$, are given. $\dot{M}$ is in terms of the present-day Sun's mass loss rate, $\dot{M_{\odot}}$. $R_{\rm{h}}$ is the calculated radius of the heliosphere. The total ionisation rate, is given at the top of the atmosphere, $\zeta^{\rm{TOA}}$, and at the surface ($P=$100 bar), $\zeta^{\rm{surf}}$.
\end{table}
\section{Results}
\label{sec:results}
In Sections\,\ref{subsec:flux} and \ref{subsec:ionisation} we present results for cosmic ray transport and the ionisation rate and ion-pair production rate of $\rm{H_2}$ by cosmic rays in the early Earth atmosphere, respectively.
\subsection{Cosmic ray differential intensity}
\label{subsec:flux}
\begin{figure}
   \centering
   \includegraphics[width=\hsize]{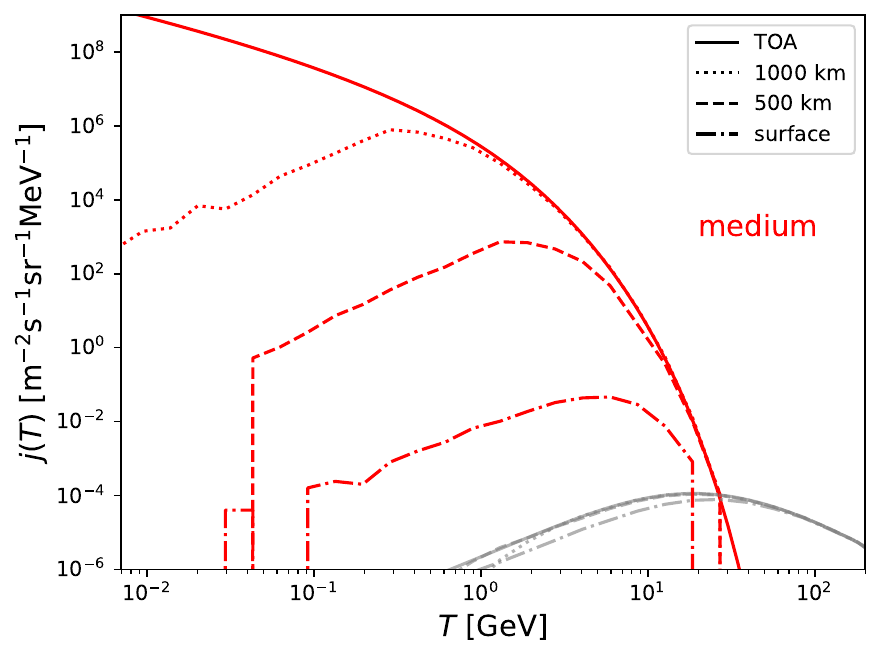}
      \caption{Differential intensity, $j(T,z)$, of solar cosmic rays (red lines) and Galactic cosmic rays (grey lines) in the early Earth atmosphere for $\Omega = 6 \Omega_{\rm{\odot}}$, plotted at different heights above and at the surface as a function of $T$. The linestyles represent these heights (solid: top of atmosphere (TOA); dotted: 1000 km; dashed: 500 km; dot-dashed: surface). Here the surface is defined as $z = 0.001$km.
              }
         \label{fig:ExampleFlux}
   \end{figure}

Fig. \ref{fig:ExampleFlux} shows $j(T,z)$ for Galactic and solar cosmic rays for a range of heights in the early Earth atmosphere, for $\Omega = 6 \Omega_{\rm{\odot}}$ (medium rotator case). Here we define the surface where $P = 100$ bar, corresponding to $z = 10^{-3} \rm{km}$. The maximum height in our model is $z = 2700 \rm{km}$ ($P = 10^{-7} \rm{bar}$), which we refer to as the top of the atmosphere. The cosmic ray spectra would be the same at any higher heights in the atmosphere not considered here. The values of $j(T,z)$ at different heights are shown in grey for Galactic cosmic rays, and in red for solar cosmic rays. In the upper part of the atmosphere ($z \gtrsim 1000\rm{km}$, corresponding to $P \lesssim 10^{-1} \rm{bar}$) where $n$ is low, only the lowest-energy solar cosmic rays show a decrease in $j(T,z)$.

The energy lost by cosmic rays with kinetic energy, $T$, over a given $dz$ in the atmosphere depends on $\sigma$ and $n$. Because $\sigma$ varies with $T$, $j(T,z)$ changes differently with height in the atmosphere for lower and higher energy cosmic rays \srr{such that} lower-energy cosmic rays ($\sim \rm{MeV}$) lose larger amounts of energy due to ionising collisions than higher-energy cosmic rays ($\sim \rm{GeV}$). This results in large decreases in $j(T,z)$ with decreasing height above the surface at low values of $T$, given a high enough $n$. For solar cosmic rays at $z \leq 1000$ km there are clear decreases in $j(T,z)$ with decreasing height, particularly at low energies ($T < 1$ GeV). For $T > 10$ GeV, $j(T,z)$ remains approximately constant until a height of 500 km. The decreases in $j(T,z)$ with increasing column depth for solar cosmic rays are greater when $\Omega$ is slower, due to the lower top-of-atmosphere values of $j(T,z)$ at high energies. For example, in Fig. \ref{fig:TOA_scr} at $T = 10 \rm{GeV}$, there is a difference of over \srr{$10^5 \rm{m^{-2}s^{-1}sr^{-1}MeV^{-1}}$} between $j(T,z)$ for $\Omega = 3.5 \Omega_{\odot}$ and \srr{$\Omega = 15 \Omega_{\odot}$}. \srr{ For solar cosmic rays there are two artificial sharp drop-offs in Fig. \ref{fig:ExampleFlux} - one above 26 GeV at $z = 500$ km (red dashed line), and another above 19 GeV at the surface (red dot-dashed line). At these heights the cosmic ray flux is so low that the number of cosmic rays in the higher energy bins goes to zero due to small sample statistics.}

The values of $j(T,z)$ for Galactic cosmic rays (grey lines in Fig. \ref{fig:ExampleFlux}), which have maximum top-of-atmosphere values at $T > 10$ GeV, remain approximately constant for $z \geq 500$ km. From 500 km (grey dashed line in Fig.\,\ref{fig:ExampleFlux}) to the surface (grey dot-dashed line) there is a small decrease in the values of $j(T,z)$ for Galactic cosmic rays. \srr{ The small change between the top-of-atmosphere and surface $j(T,z)$ of Galactic cosmic rays is due to the fact that we only include cosmic ray energy losses due to ionisation. Since the Galactic cosmic ray spectrum peaks at high energies, where $\sigma$ is smaller than at lower energies, the ionisation energy losses are small and the spectrum remains approximately constant with height. A larger change in $j(T,z)$ between the top of the atmosphere and the surface would be expected if additional energy loss mechanisms such as those due to pion production were included \citep[see e.g. Fig. 7][]{padovani-2009}.} For $\Omega = 3.5 \Omega_{\rm{\odot}}$ and \srr{$\Omega = 15 \Omega_{\rm{\odot}}$}, not shown here, the results for $j(T,z)$  \srr{show similar behaviour} for Galactic cosmic rays. 

The relevant quantities for chemical modelling to account for cosmic ray effects on the early Earth atmosphere are $\zeta$ (Eq. \ref{eq:zeta}) and the ion-pair production rate, \srr{$Q(z) = n(z) \zeta (z)$}. In Section \ref{subsec:ionisation} we present \srr{$\zeta(z)$ and $Q(z)$} using $j(T,z)$.
 
\subsection{$\rm{H_2}$ ionisation and ion-pair production rates}
\label{subsec:ionisation}
\begin{figure}
   \centering
   \includegraphics[width=\hsize]{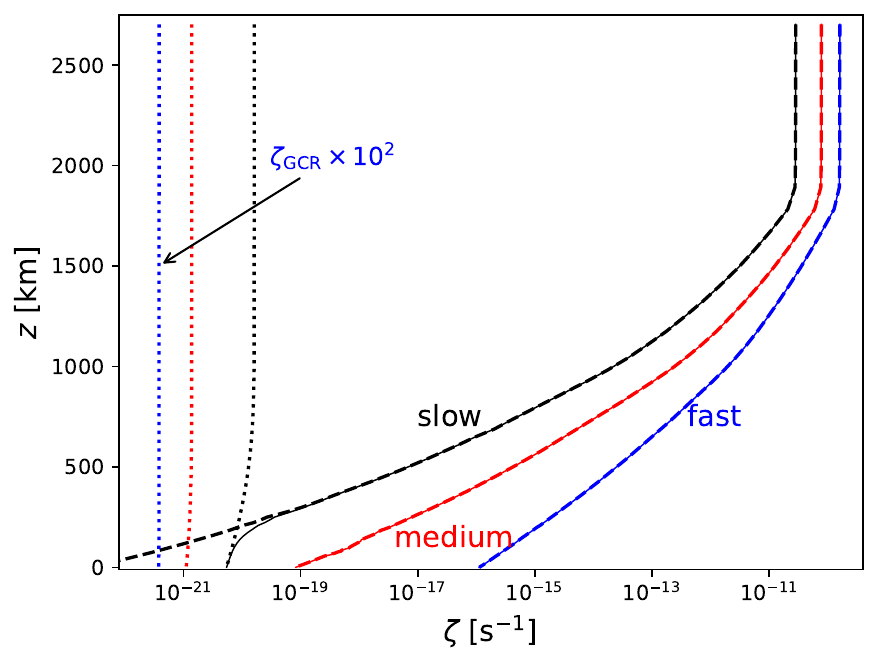}
      \caption{Ionisation rate, $\zeta$, as a function of height in the early Earth atmosphere at 200 Myr due to Galactic and solar cosmic rays. Dotted lines show $\zeta$ due to Galactic cosmic rays only, dashed lines show $\zeta$ due to solar cosmic rays only, solid lines show the total $\zeta$ from Galactic cosmic rays and solar cosmic rays combined. The $\zeta$ values for the slow, medium and fast solar rotation rates are shown by black, red and blue lines, respectively. For the fast solar rotation rate, $\zeta_{\rm{GCR}}$ has been multiplied by \srr{$10^2$ }to allow $\zeta_{\rm{GCR}}$ and $\zeta_{\rm{SCR}}$ for all rotation rates to be shown within a reasonable range on the same axes.
              }
         \label{fig:ZetaPlot}
   \end{figure}
Fig. \ref{fig:ZetaPlot} shows $\zeta$ \srr{(using Eq. \ref{eq:zeta})} for Galactic \srr{($\zeta_{\rm{GCR}}$, dotted lines)} and solar cosmic rays \srr{($\zeta_{\rm{SCR}}$, dashed lines)} as a function of height in the post-impact early Earth atmosphere at 200 Myr, for the three values of $\Omega$ given in Table \ref{table:rotation-rates}. The total $\zeta = \zeta_{\rm{GCR}} + \zeta_{\rm{SCR}}$ is represented by thin solid lines. For each type of cosmic ray, the values of $\zeta$ for the slow, medium and fast rotator cases are shown by the black, red and blue lines, respectively. 

 We focus on $\zeta$ at the surface, $\zeta^{\rm{surf}}$, to understand the effects of cosmic rays where life emerged. For the fast rotator case, taking $10^{-18}\rm{s}^{-1}$ as an estimate for an ionisation rate high enough to have an effect on the chemistry, \srr{$\zeta^{\rm{surf}} \sim 10^{-16}\rm{s}^{-1}$} {resulting from solar cosmic rays} is likely high enough. In the slow or medium rotator cases $\zeta^{\rm{surf}}$ (given in Table\,\ref{table:results}) is not high enough to be important for chemistry at the surface. \srr{Thus,} we find that for solar cosmic rays to be important for chemistry at the surface of the post-impact early Earth, the Sun must have \srr{been a fast rotator}. We consider the suppression of the top-of-atmosphere cosmic ray spectra by a planetary magnetic field in Appendix \ref{subsec:bfield}. We find that the suppression does not result in changes to $\zeta^{\rm{surf}}$ at 200 Myr.
 
 On the other hand, it is possible that prebiotic molecules produced in the early Earth atmosphere could have been dissolved in atmospheric water droplets and precipitated to the surface \citep{benner-2020}. In this case it is useful to consider $\zeta$ throughout the early Earth atmosphere. The total values of $\zeta$ at the top of the early Earth atmosphere, $\zeta^{\rm{TOA}}$, are given in Table\,\ref{table:results}. At the top of the atmosphere $\zeta_{\rm{SCR}} > \zeta_{\rm{GCR}}$ for all values of $\Omega$ such that $\zeta \simeq \zeta_{\rm{SCR}}$. For example, for $\Omega =6 \Omega_{\rm{\odot}}$, $\zeta_{\rm{SCR}} = 8\times 10^{-11}\rm{s^{-1}}$ at the top of the atmosphere, over 10 orders of magnitude higher than $\zeta_{\rm{GCR}} = 10^{-21}\rm{s^{-1}}$. 
 
The top-of-atmosphere values of both $\zeta_{\rm{SCR}}$ and $\zeta_{\rm{GCR}}$ vary greatly depending on the value of $\Omega$ used. In the fast rotator case, $\zeta_{\rm{SCR}}=2\times 10^{-10}\rm{s^{-1}}$ at the top of the atmosphere, nearly an order of magnitude greater than $\zeta_{\rm{SCR}}=3\times 10^{-11}\rm{s^{-1}}$ for the slow rotator. This reflects the top-of-atmosphere solar cosmic ray spectra which show higher $j(T,z)$ for the fast rotator than for the slow rotator. The changes in $\zeta$ as a function of height are similar to the changes in $j(T,z)$ described in Section \ref{subsec:flux}: $\zeta_{\rm{GCR}}$ is almost constant with height and $\zeta_{\rm{SCR}}$ (dashed lines in Fig.\,\ref{fig:ZetaPlot}) is large and constant in the upper atmosphere, but decreases rapidly below $z=1700$ km. The decrease in $\zeta_{\rm{SCR}}$ for $z < 1700$ km is not as rapid in the fast rotator case (blue dashed line in Fig.\,\ref{fig:ZetaPlot}) as it is in the medium (red dashed line) and slow rotator (black dashed line) cases. This is because $p_{\rm{max}}$ is greater and the decrease in $j(T,z)$ with decreasing $z$ at low energies is more gradual for faster $\Omega$. For the fast rotator, \srr{$\zeta_{\rm{SCR}}^{\rm{surf}} = 1\times 10^{-16}\rm{s^{-1}}$}. For the slow rotator, $\zeta_{\rm{SCR}}^{\rm{surf}} = 2\times 10^{-23}\rm{s^{-1}}$ is lower than $\zeta_{\rm{GCR}} = 5\times 10^{-21}\rm{s^{-1}}$. $\zeta$ is indistinguishable from $\zeta_{\rm{SCR}}$ in Fig.\,\ref{fig:ZetaPlot}, except for $z \lesssim 300$ km for the slow rotator, where $\zeta_{\rm{SCR}} < \zeta_{\rm{GCR}}$. 
 This suggests that solar cosmic rays have a greater effect on chemistry throughout the early Earth atmosphere than Galactic cosmic rays. 
 
\begin{figure}
   \centering
   \includegraphics[width=\hsize]{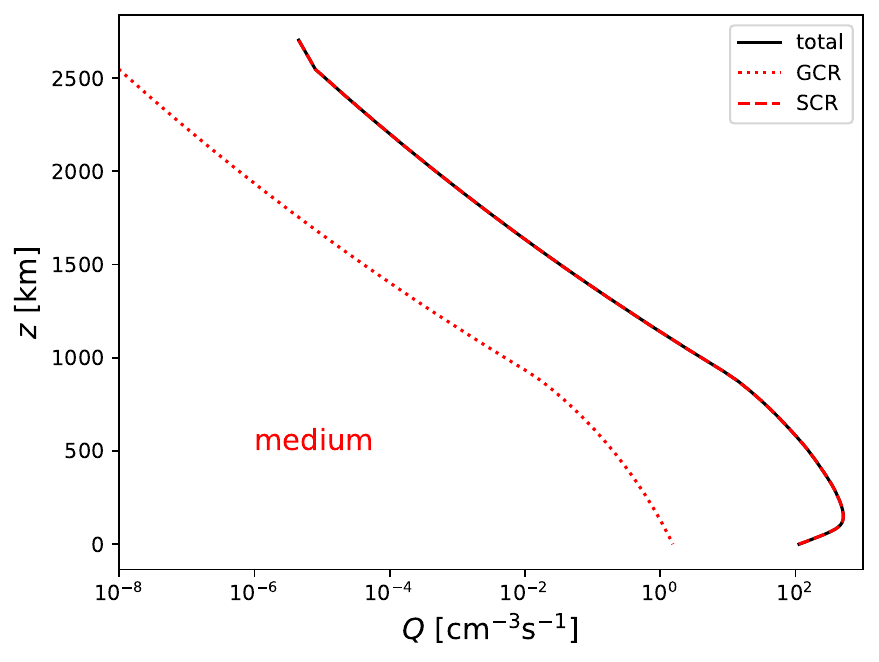}
      \caption{The ion-pair production rate, $Q$, as a function of height in the early Earth atmosphere at 200 Myr for $\Omega = 6\Omega_{\odot}$. The solid black line shows the total $Q$ value: the sum of $Q$ for Galactic cosmic rays (`GCR', red dotted line) and for solar cosmic rays (`SCR', red dashed line).
              }
         \label{fig:qmed}
   \end{figure}

 Fig. \ref{fig:qmed} shows $Q(z)$ in the early Earth atmosphere for both Galactic ($Q_{\rm{GCR}}$, red dotted line) and solar ($Q_{\rm{SCR}}$, red dashed line) cosmic rays for $\Omega = 6\Omega_{\rm{\odot}}$ (medium rotator case). The total $Q = Q_{\rm{GCR}} + Q_{\rm{SCR}}$ as a function of height is shown by the solid black line. \srr{For comparison, for the present-day Earth, balloon measurements by \citet{neher-1971} for $16^{\rm{th}}$ April 1965 covering $6\, \rm{km}<\textit{z}<32\,\rm{km}$ show a maximum of $Q = 37.48 \,\rm{cm^{-3}s^{-1}}$ at $z = 14\, \rm{km}$.} Similar to $\zeta$, $Q_{\rm{SCR}} \gg Q_{\rm{GCR}}$ for $\Omega = 6\Omega_{\rm{\odot}}$ at all heights and the total $Q$ is indistinguishable from $Q_{\rm{SCR}}$. The maximum value of $Q_{\rm{SCR}}$ is at $z = 150$ km ($P = 50$ bar). \srr{If additional energy losses such as those due to pion production were included $\zeta_{\rm{GCR}}$ would decrease with decreasing height and lead to the maximum of $Q_{\rm{GCR}}$ occurring at some $z$ in the atmosphere and not in the soil as our results currently suggest. Air shower models such as AtRIS \citep{banjac-2019} can be used in the future to include these additional energy losses.}

 The values of $Q_{\rm{SCR}}$ and $Q_{\rm{GCR}}$ presented in Fig.\,\ref{fig:qmed} indicate that solar cosmic rays are more important for chemistry than Galactic cosmic rays at all heights in the early Earth atmosphere for $\Omega = 6\Omega_{\rm{\odot}}$. Similarly, $Q_{\rm{SCR}} > Q_{\rm{GCR}}$ throughout the atmosphere for \srr{$\Omega = 15\Omega_{\rm{\odot}}$}. For $\Omega = 3.5\Omega_{\rm{\odot}}$, $Q_{\rm{GCR}} > Q_{\rm{SCR}}$ for $z \leq 300$ km ($P \geq 20$ bar), where $\zeta < 10^{19} \rm{s}^{-1}$ appears too low to have a significant effect on the atmospheric chemistry. 

\section{Discussion}
\label{sec:discussion}
\srr{\subsection{Variations of the Galactic cosmic ray spectrum}}
    \label{subsec:vary-lis}
    \begin{figure}
        \centering
        \includegraphics[width=\hsize]{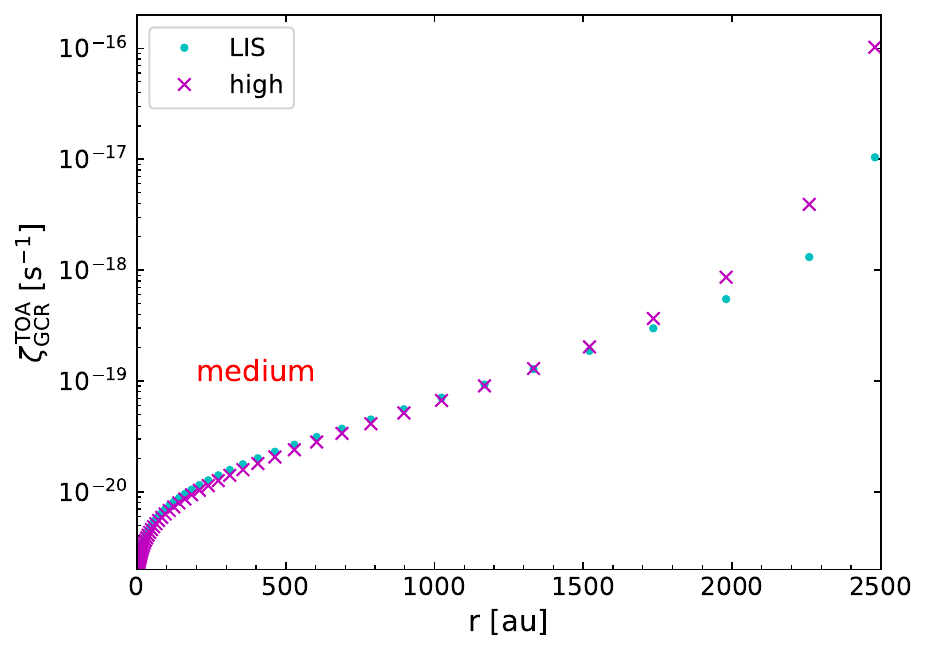}
        \caption{\srr{$\zeta_{\rm{GCR}}^{\rm{TOA}}$}, calculated using top-of-atmosphere spectra at varying distances from the Sun in the medium rotator case, $\Omega = 6\Omega_{\rm{\odot}}$. The circles (crosses) represent \srr{$\zeta_{\rm{GCR}}^{\rm{TOA}}$} calculated using top-of-atmosphere spectra derived from $j_{\rm{LIS}}$ ($j_{\rm{LIS}}^{\rm{high}}$).}
        \label{fig:vary-lis}
    \end{figure}
    The Galactic cosmic ray results in Sections \ref{subsec:flux} - \ref{subsec:ionisation} are based on the assumption that the LIS is approximately constant on Gyr time scales \citep[ignoring short-term effects related to the solar cycle,][]{potgieter-2013}. \srr{ However, there are a number of factors that can change the Galactic cosmic ray spectrum both spatially and temporally in the Galaxy. Firstly, the Galactic cosmic ray spectrum is different inside and outside the Galaxy's spiral arms. The model presented by \citet{busching-2008} showed approximately a factor of 2 increase in cosmic ray fluxes inside the spiral arms compared to the inter-arm regions due to cosmic ray sources, assumed to be supernova remnants, in the spiral arms. This is in agreement with $\gamma$-ray observations - \citet{aharonian-2020} found an increase by a factor of 2 - 4 in the flux of cosmic rays with energies $> 10$ GeV within 100 pc of supernova remnants, compared to the homogeneous "sea" of cosmic rays observed throughout the Galaxy. The rate of supernova remnants, which is dependent on the star formation rate in the Galaxy, would also have influenced the Galactic cosmic ray spectrum on Gyr timescales. Assuming that the Galactic cosmic ray flux is proportional to the star formation rate in the Galaxy, \citet{svensmark-2006} indicated a factor of 1.4 increase in the Galactic cosmic ray flux at 200 Myr compared to the present day due to the increased supernova remnant/star formation rate.

     The presence of young low-mass stars in star-forming regions will also influence the Galactic cosmic ray spectrum, mainly at MeV energies.} Here we investigate the effect of varying the LIS \srr{ to reflect this influence} on the top-of-atmosphere values of $j(T)$ and $\zeta_{\rm{GCR}}(z)$ for the early Earth. Because the majority of stars form within star clusters \citep{arakawa-2023}, it is likely that there were many young low-mass stars close to the Sun at 200 Myr. Taking cosmic rays accelerated at these young low-mass stars into account \citep[e.g. ][]{padovani-2016}, $j(T)$ in the LIS \srr{was} likely higher than at present, particularly at $\sim$MeV energies \srr{\citep[shown by the dot-dashed grey line in Fig. \ref{fig:TOA_gcr} as $j_{\rm{LIS}}^{\rm{high}}$][]{padovani-2018}}. $j_{\rm{LIS}}^{\rm{high}}(T)$ has higher values than $j_{\rm{LIS}}$ (Eq. \ref{eqn:vos}) at low energies to reproduce the cosmic ray ionisation rates observed in molecular clouds, that cannot be explained by a simple extrapolation of \textit{Voyager 1} data to low energies \srr{and are likely due to increased cosmic ray production from nearby young \srr{low-mass} stars}. This spectrum, labelled $\mathcal{H}$ in \citet{padovani-2018}, is given by:
    \begin{equation}
    \label{eqn:lis-pad}
        j_{\rm{LIS}}^{\rm{high}}(T) = 2.4 \times 10^{15} \frac{T^{-0.8}}{(T + T_0)^{1.9}} \rm{eV^{-1}s^{-1}cm^{-2}sr^{-1}}\,,
    \end{equation}
where $T_0 = 6.50 \times 10^8$ eV and $T$ is in units of eV.  

Using the \srr{same solar wind properties (Table \ref{table:results}) and cosmic ray transport model as in Section \ref{subsec:solar-transport}}, we have produced new Galactic cosmic ray spectra at different orbital distances \srr{using $j_{\rm{LIS}}^{\rm{high}}(T)$}. Ignoring the effect of a planetary magnetic field, these spectra would be the hypothetical top-of-atmosphere spectra at these orbital distances. Fig. \ref{fig:vary-lis} shows $\zeta_{\rm{GCR}}$ at the top of the early Earth atmosphere, $\zeta_{\rm{GCR}}^{\rm{TOA}}$, as a function of orbital distance, $r$, calculated using the hypothetical top-of-atmosphere spectra. We compare $\zeta_{\rm{GCR}}^{\rm{TOA}}$ resulting from $j_{\rm{LIS}}$ (circles) and $j_{\rm{LIS}}^{\rm{high}}$ (crosses), for the medium rotator case as an example. 

We find at 1~au, where the low-energy part of the Galactic cosmic ray spectrum is suppressed due to the solar wind, that the spectrum is unchanged by varying the LIS. Thus, $\zeta_{\rm{GCR}}^{\rm{TOA}}$ for the medium rotation rate is the same ($\zeta_{\rm{GCR}}^{\rm{TOA}} = 10^{-21}\rm{s^{-1}}$) for both $j_{\rm{LIS}}^{\rm{high}}$ and $j_{\rm{LIS}}$, indicating that an increase in $j_{\rm{LIS}}$ at low energies would not have a significant effect on $\zeta_{\rm{GCR}}$ in Earth's atmosphere at 200 Myr for the scenario considered here. The low-energy Galactic cosmic rays at 1~au are not in fact related to the low-energy LIS cosmic rays. Instead, $j$ at low energies at 1~au depends on $j$ in the LIS at higher energies. The high-energy Galactic cosmic rays propagating through the \srr{Solar System} experience energy losses when doing work against the solar wind and contribute to $j$ at lower energies at 1au. 
    
    More generally, for $r < 1300$au, the top-of-atmosphere spectra resulting from $j_{\rm{LIS}}$ and $j_{\rm{LIS}}^{\rm{high}}$ are similar -  the difference between $\zeta_{\rm{GCR}}^{\rm{TOA}}$ calculated using the two top-of-atmosphere spectra does not exceed $10 \%$. However, Fig. \ref{fig:vary-lis} shows that $\zeta_{\rm{GCR}}^{\rm{TOA}}$ derived from $j_{\rm{LIS}}(T)$ is slightly greater than $\zeta_{\rm{GCR}}^{\rm{TOA}}$ derived from $j_{\rm{LIS}}^{\rm{high}}(T)$ for $r < 1300$au. The small differences  are due to the slightly lower values of $j_{\rm{LIS}}^{\rm{high}}(T)$ compared to $j_{\rm{LIS}}(T)$ at high energies, visible for $T > 5$ GeV in Fig. \ref{fig:TOA_gcr}.
    
    However, for hypothetical planets orbiting \srr{at $r \geqslant 1300$au}, the effect on $\zeta_{\rm{GCR}}^{\rm{TOA}}$ of varying the LIS of Galactic cosmic rays is greater. Compared to $j_{\rm{LIS}}$, \srr{$j_{\rm{LIS}}^{\rm{high}}(T)$ is larger} for low-energy Galactic cosmic rays. At large values of $r$ the suppression of Galactic cosmic rays is not extreme enough to make this difference insignificant in the modelled top-of-atmosphere spectra. The higher intensity of low-energy cosmic rays with large ionisation cross-sections results in a higher $\zeta_{\rm{GCR}}^{\rm{TOA}}$. At 2300~au, $\zeta_{\rm{GCR}}^{\rm{TOA}}$ calculated using the top-of-atmosphere spectrum resulting from $j_{\rm{LIS}}^{\rm{high}}(T)$ is a factor of 3 larger than the value calculated using the spectrum resulting from $j_{\rm{LIS}}$. Outside of the \srr{Solar System} $\zeta_{\rm{LIS}}^{\rm{high}} = 10 ^{-16}\rm{s}^{-1}$, which is an order of magnitude greater than  $\zeta_{\rm{LIS}} = 10 ^{-17}\rm{s}^{-1}$. \srr{For the slow rotator case the effect of changing the LIS is seen at smaller orbital distances. For $r < 400$au, $\zeta_{\rm{GCR}}^{\rm{TOA}}$ calculated using both top-of-atmosphere spectra are similar. The factor of 3 increase in $\zeta_{\rm{GCR}}^{\rm{TOA}}$ when using $j_{\rm{LIS}}^{\rm{high}}(T)$ seen at 2300~au for the medium rotator case occurs at $\sim$1400~au for the slow rotator case.}
    
    We find that assuming a LIS which changes at lower energies over Gyr time scales has a minimal impact on the calculated ionisation rate at 1~au. To affect $\zeta_{\rm{GCR}}^{\rm{TOA}}$ at 1~au, the LIS would need to have higher $j$ values compared to $j_{\rm{LIS}}$ at higher energies ($T > 1$ GeV). This could be caused by the acceleration of Galactic cosmic rays by a nearby supernova remnant. At larger orbital distances where Galactic cosmic rays are subject to less extreme suppression the LIS used has a greater effect on the resulting ionisation rate at the top of the atmosphere. When comparing $\zeta_{\rm{GCR}}$ and $\zeta_{\rm{SCR}}$ for exoplanets orbiting at large distances from their host stars, it will be important to consider the Galactic cosmic ray spectra resulting from different possible LIS. The stellar wind properties of the host star will also play an important role in determining the suppression of the Galactic cosmic ray spectrum and the orbital distance at which a change in the LIS affects $\zeta_{\rm{GCR}}^{\rm{TOA}}$. For example, keeping the level of turbulence constant, an increased stellar wind velocity increases the suppression of the Galactic cosmic ray spectrum and a change in the LIS produces a change in $\zeta_{\rm{GCR}}^{\rm{TOA}}$ at a greater orbital distance. Investigating ionisation due to cosmic rays in exoplanet atmospheres orbiting at large distances from a host star with a stellar wind similar to the young Sun requires knowledge of the LIS.
    
\subsection{\srr{Importance for chemistry and exoplanet atmospheres}}
\label{subsec:discussion-chem}
The influence of cosmic rays on the chemistry of (exo)planetary atmospheres is relevant both for the search for signatures of life on exoplanets and the study of the origin of life on Earth. \citet{drl-2023} modelled the transport of stellar and Galactic cosmic rays through the $\rm{H_2}$-dominated atmosphere of a gas giant with an M dwarf host star. The result shown in Fig. 6 of \citet{drl-2023} is that stellar cosmic rays dominate until the atmospheric pressure exceeds 10 bar, and $Q_{\rm{GCR}}(P = 100 \rm{bar}) \simeq 1 \rm{cm^{-3}s^{-1}}$. This result is similar to the slow rotator case for the early Earth atmosphere discussed here, where $Q_{\rm{SCR}} > Q_{\rm{GCR}}$ for $P < 20 \rm{bar}$ and $Q_{\rm{GCR}}(P = 100 \rm{bar}) = 10 \rm{cm^{-3}s^{-1}}$. \srr{ Modelling of M dwarf exoplanet systems \citep[e.g.][]{herbst-2020,mesquita-2021,mesquita-2022} has focused on the 1D transport and role of cosmic rays in these systems.} Studying exoplanets observable with JWST transmission spectroscopy can provide more information on the cosmic ray effects on atmospheres similar to the early Earth atmosphere and improve our understanding of the origin of life.

\citet{rimmer-2016} presented an ion-neutral chemical model, used to study the chemical reactions in a selection of atmospheres including an early Earth atmosphere dominated by $\rm{CO_2}$ and $\rm{N_2}$. Along with chemistry driven by UV light and lightning, this model \srr{took} into account chemistry driven by cosmic ray ionisation. The cosmic ray ionisation rate in the atmosphere is therefore an important input. Chemical modelling \citep[e.g.][]{helling-2019} shows that cosmic ray ionisation is important for producing molecules such as $\rm{H_3O^+}$ and enhancing \srr{prebiotic molecule production} \citep{barth-2021}. Our results can be used in chemical models to improve our understanding of prebiotic chemistry in the post-impact early Earth scenario investigated here. \srr{In the future, to use our results for $\zeta$ in chemical modelling, it would be important for the chemical models to include all of the relevant molecular abundances in the atmosphere, such as those given in Table 1 of \citet{zahnle-2020} for $\rm{CO_2}$, $\rm{CH_4}$ and $\rm{NH_3}$ in addition to the $\rm{H_2}$ included here in the cosmic ray model.}

The effects of cosmic rays on the chemistry of $\rm{H_2}$-dominated atmospheres are relevant in the search for the chemical signatures of life on exoplanets \srr{ \citep[see e.g.][for a recent review]{herbst-2022}}. \srr{The upper atmosphere becomes relevant when considering the applications of cosmic ray transport modelling beyond Earth. Exoplanets observable by JWST \citep{dyrek-2024,gardner-2006,rigby-2023} include warm Neptune planets with $\rm{H_2}$-dominated atmospheres \citep[e.g.][]{drl-2023}. The effects of cosmic rays on the chemistry of the upper atmospheres of these exoplanets are important for interpreting transmission spectroscopy observations probing low pressures \citep{madhusudhan-2019}.} Hycean worlds - planets with potentially habitable ocean surfaces underneath $\rm{H_2}$-rich atmospheres - are promising candidates for the detection of biosignatures using JWST NIRSpec and NIRISS transmission spectroscopy \citep{madhusudhan-2021}. There has been much interest in JWST observations of the exoplanet K2-18 b, which has been interpreted as a Hycean world \citep{madhusudhan-2023} and later as a gas giant with no habitable surface \citep{wogan-2024}. The interpretation of observations of exoplanets orbiting active stars using chemical models may be affected by the high stellar cosmic ray ionisation expected in the exoplanet atmospheres. 

\srr{Additionally, the} $\rm{H_2}$-dominated atmospheres of exoplanets in diverse environments act as "laboratories" to search for prebiotic chemistry \citep{rimmer-2016} where observations of the early Earth atmosphere itself are not possible. \citet{claringbold-2023} found that prebiotic molecules such as cyanoacetylene and formaldehyde are well suited for detection in $\rm{H_2}$-rich exoplanet atmospheres\srr{, such as Hycean worlds and post-impact planets,} using the JWST NIRSpec and NIRISS instruments. The results presented by \citet{claringbold-2023} emphasise the strong capabilities of JWST transmission spectroscopy observations of $\rm{H_2}$-dominated exoplanet atmospheres to advance the study of the origin of life.

\subsection{\srr{Additional cosmic ray effects}}
\label{subsec:pions-etc}
The transport model described in this paper takes into account the energy lost by cosmic rays through ionising collisions. For the energy range ($10^{-2}\rm{GeV}\lesssim \textit{T} \lesssim 10^{2}\rm{GeV}$) discussed in this paper the ionisation cross section is larger for lower-energy cosmic rays and the higher-energy cosmic rays reach deeper into the atmosphere before losing energy to ionisation collisions. The model adopted here does not take into account the additional energy lost to pion production by cosmic ray protons with energies above $\sim 1$ GeV \citep[see Fig. 7 of][]{padovani-2009}. Pion production is only a first step in a cascade or air shower, where secondary particles are produced and go on to have additional effects on the chemistry in the atmosphere. The particle transport model presented by \citet{herbst-2019} includes secondary particle effects on chemistry in the present-day Earth's atmosphere. Above $\sim 1$ GeV the energy lost to pion production increases with increasing cosmic ray proton energy \citep[e.g.][]{padovani-2009}, suggesting that the differential intensity in the post-impact early Earth atmosphere presented here may be an overestimation for higher-energy cosmic rays reaching the surface. However, experiments studying different early Earth atmosphere compositions including CO, $\rm{CO_2}$, $\rm{N_2}$ and $\rm{H_2O}$ show that the secondary electrons produced by these high-energy cosmic rays are also interesting because they can break $\rm{N_2}$ bonds and lead to increased production of HCN \citep{airapetian-2020}. 

Other secondary cosmic ray particles such as muons interact with prebiotic molecules and can produce additional changes to these molecules. \citet{globus-2020} suggested that interactions between spin-polarised cosmic ray muons and biological molecules is one explanation for the chirality of DNA and RNA helices. Future calculations of the secondary particle fluxes associated with our cosmic ray proton spectra in the early Earth atmosphere would provide further insight into the role of cosmic rays in the origin of life.

\section{Conclusions}
\label{sec:conclusions}
In this paper we have modelled the transport of Galactic and solar cosmic rays in a transient $\rm{H_2}$-dominated atmosphere after an ocean-vaporising impact, known as the post-impact early Earth atmosphere. This atmosphere, low in oxygen, presents a favourable environment for the formation of prebiotic molecules, if a source of ionisation is also present. Our aim is to compare ionisation by Galactic and solar cosmic rays in the early Earth atmosphere, particularly at the surface where life likely formed. To achieve this we have calculated the differential intensity of Galactic and solar cosmic rays separately as a function of height in the atmosphere. We have calculated the resulting $\rm{H_2}$ ionisation rates as a function of atmospheric height.
The Galactic and solar cosmic ray spectra arriving at the Earth are determined by the properties of the Sun and solar wind, which both depend on the Sun's rotation rate. For the post-impact early Earth scenario at 200 Myr, the Sun's rotation rate is unknown. To represent the range of possible rotation rates of the young Sun, we have modelled the cosmic ray spectra in three scenarios - slow ($\Omega = 3.5 \Omega_{\rm{\odot}}$), medium ($\Omega = 6 \Omega_{\rm{\odot}}$) and fast \srr{($\Omega = 15 \Omega_{\rm{\odot}}$)} rotator cases, spanning the range of observed rotation rates \srr{for low-mass stars}. 

For each of the three assumed solar rotation rates, the differential intensity of Galactic cosmic rays at the top of the early Earth atmosphere is at its maximum at high energies ($T > 1 \rm{GeV}$). These high-energy cosmic rays lose little energy to ionisation in the atmosphere, resulting in Galactic cosmic ray spectra at the surface which are similar to the top-of-atmosphere spectra. In contrast, solar cosmic rays have maximum top-of-atmosphere differential intensities at low energies ($T < 1 \rm {GeV}$). At low energies the differential intensity decreases with decreasing height, producing solar cosmic ray spectra which differ in both overall differential intensity and shape compared to the top-of-atmosphere spectra.

We find that for each of the three solar rotation rates, the ionisation rate throughout the atmosphere, except for $z \lesssim 300 \rm{km}$ for the slow rotator case, is dominated by solar cosmic rays. However, only in the fast rotator case is the ionisation rate at the surface (\srr{$\zeta^{\rm{surf}} \sim 10^{-16}\rm{s}^{-1}$}) high enough to be of interest for chemistry. The ion-pair production rate reaches its maximum in the lower part of atmosphere, $\sim 100-300$km above the surface. For all three solar rotation rates, the ion-pair production rate at its maximum is also dominated by solar cosmic rays. Galactic cosmic rays become dominant - but not of interest for chemistry -  only for the slow rotator case, for $z < 300$km.

We consider the effect of the Earth's magnetic field on the top-of-atmosphere cosmic ray spectra. Low-energy cosmic rays are affected by Earth's magnetic field and are deflected towards the poles. This deflection particularly affects lower-energy cosmic rays below a cutoff energy. We calculate the cutoff energy imposed on the top-of-atmosphere cosmic ray spectra by the early Earth's magnetic field to be $T_{\rm{C}}=2.56\rm{GeV}$. We modify the top-of atmosphere spectra to account for the deflection of cosmic rays with energies below this cutoff. High-energy ($T > 2.5 \rm{GeV}$) cosmic rays are not deflected. At the surface the ionisation rate is not related to the low-energy part of the top-of-atmosphere spectrum, and is not affected by the suppression of this spectrum at low energies. 

These results assume that the spectrum of Galactic cosmic rays outside the \srr{Solar System} is constant over Gyr timescales. To account for the possibility of young low-mass stars near the \srr{Solar System} at 200 Myr increasing the flux of low-energy ($T<1 \rm{GeV}$) cosmic rays compared to today, we have compared the top-of-atmosphere ionisation rate by Galactic cosmic rays using the present-day local interstellar spectrum and a spectrum which has increased cosmic ray flux at low energies (more than a factor of 10 larger at $10^{-2}$GeV). Interestingly, at 1~au this change to the low-energy end of the LIS has no effect on the Galactic cosmic ray ionisation rate. At larger orbital distances where the suppression of the LIS by the solar wind is less severe the difference becomes more significant - for $\Omega = 6 \Omega_{\rm{\odot}}$ at 2300 au the ionisation rate is a factor of 3 larger when using the higher LIS. For applications of cosmic ray transport modelling to exoplanet systems with a stellar wind similar to the young Sun and a fixed level of turbulence, the use of a different LIS is likely to affect the resulting top-of-atmosphere cosmic ray spectra only at large orbital distances. 

\srr{The fact that the top-of-atmosphere Galactic cosmic ray spectra at 1\,au resulted in extremely low ionisation rates depends on the turbulence and ISM properties adopted. 3D particle drift effects that are not captured in our 1D transport model could lead to an increase in the top-of-atmosphere Galactic cosmic ray fluxes and warrants further 3D transport modelling in the future. At the same time,} we find that in the post-impact early Earth atmosphere, solar cosmic rays are an important source of ionisation that should be taken into account for planetary chemical modelling in the future. Solar cosmic rays dominate the ionisation rate throughout the planet's atmosphere. Depending on the rotation rate of the young Sun, we find that this ionisation can be high enough to significantly alter the chemistry down on the Earth's surface.  

\begin{acknowledgements}
    \srr{We thank the anonymous reviewer for their constructive comments which improved the paper.} SRR and DRL would like to acknowledge that this publication has emanated from research conducted with the financial support of Taighde {\'E}ireann – Research Ireland under Grant number 21/PATH-S/9339. SRR would like to thank Andrew Taylor and Tom Ray for helpful discussions which improved the paper.
\end{acknowledgements}

\bibliographystyle{aa}
\bibliography{bib-file}

\begin{thebibliography}{80}
\expandafter\ifx\csname natexlab\endcsname\relax\def\natexlab#1{#1}\fi

\bibitem[{Aguilar {et~al.}(2002)Aguilar, Alcaraz, Allaby, Alpat, Ambrosi, Anderhub, Ao, Arefiev, Azzarello, Babucci, Baldini, Basile, Barancourt, Barao, Barbier, Barreira, Battiston, Becker, Becker, Bellagamba, Béné, Berdugo, Berges, Bertucci, Biland, Bizzaglia, Blasko, Boella, Boschini, Bourquin, Brocco, Bruni, Buénerd, Burger, Burger, Cai, Camps, Cannarsa, Capell, Casadei, Casaus, Castellini, Cecchi, Chang, Chen, Chen, Chen, Chernoplekov, Chiueh, Cho, Choi, Choi, Chuang, Cindolo, Commichau, Contin, Cortina-Gil, Cristinziani, {da Cunha}, Dai, Delgado, Deus, Dinu, Djambazov, D'Antone, Dong, Emonet, Engelberg, Eppling, Eronen, Esposito, Extermann, Favier, Fiandrini, Fisher, Fluegge, Fouque, Galaktionov, Gervasi, Giusti, Grandi, Grimms, Gu, Hangarter, Hasan, Hermel, Hofer, Huang, Hungerford, Ionica, Ionica, Jongmanns, Karlamaa, Karpinski, Kenney, Kenny, Kim, Kim, Kim, Kim, Klimentov, Kossakowski, Koutsenko, Kraeber, Laborie, Laitinen, Lamanna, Lanciotti, Laurenti, Lebedev, Lechanoine-Leluc, Lee, Lee, Levi,
  Levtchenko, Liu, Liu, Lopes, Lu, Lu, Lübelsmeyer, Luckey, Lustermann, Maña, Margotti, Mayet, McNeil, Meillon, Menichelli, Mihul, Mourao, Mujunen, Palmonari, Papi, Park, Park, Pauluzzi, Pauss, Perrin, Pesci, Pevsner, Pimenta, Plyaskin, Pojidaev, Pohl, Postolache, Produit, Rancoita, Rapin, Raupach, Ren, Ren, Ribordy, Richeux, Riihonen, Ritakari, Ro, Roeser, Rossin, Sagdeev, Santos, Sartorelli, Sbarra, Schael, {Schultz von Dratzig}, Schwering, Scolieri, Seo, Shin, Shoutko, Shoumilov, Siedling, Son, Song, Steuer, Sun, Suter, Tang, Ting, Ting, Tornikoski, Torsti, Trümper, Ulbricht, Urpo, Valtonen, Vandenhirtz, Velcea, Velikhov, Verlaat, Vetlitsky, Vezzu, Vialle, Viertel, Vité, Gunten, Wicki, Wallraff, Wang, Wang, Wang, Wiik, Williams, Wu, Xia, Yan, Yan, Yang, Yang, Yang, Ye, Yeh, Xu, Zhang, Zhang, Zhao, Zhu, Zhu, Zhuang, Zichichi, Zimmermann, \& Zuccon}]{aguilar-2002}
Aguilar, M., Alcaraz, J., Allaby, J., {et~al.} 2002, Physics Reports, 366, 331

\bibitem[{Aharonian {et~al.}(2020)Aharonian, Peron, Yang, Casanova, \& Zanin}]{aharonian-2020}
Aharonian, F., Peron, G., Yang, R., Casanova, S., \& Zanin, R. 2020, Phys. Rev. D, 101, 083018

\bibitem[{{Airapetian} {et~al.}(2020){Airapetian}, {Barnes}, {Cohen}, {Collinson}, {Danchi}, {Dong}, {Del Genio}, {France}, {Garcia-Sage}, {Glocer}, {Gopalswamy}, {Grenfell}, {Gronoff}, {G{\"u}del}, {Herbst}, {Henning}, {Jackman}, {Jin}, {Johnstone}, {Kaltenegger}, {Kay}, {Kobayashi}, {Kuang}, {Li}, {Lynch}, {L{\"u}ftinger}, {Luhmann}, {Maehara}, {Mlynczak}, {Notsu}, {Osten}, {Ramirez}, {Rugheimer}, {Scheucher}, {Schlieder}, {Shibata}, {Sousa-Silva}, {Stamenkovi{\'c}}, {Strangeway}, {Usmanov}, {Vergados}, {Verkhoglyadova}, {Vidotto}, {Voytek}, {Way}, {Zank}, \& {Yamashiki}}]{airapetian-2020}
{Airapetian}, V.~S., {Barnes}, R., {Cohen}, O., {et~al.} 2020, International Journal of Astrobiology, 19, 136

\bibitem[{{Airapetian} {et~al.}(2016){Airapetian}, {Glocer}, {Gronoff}, {H{\'e}brard}, \& {Danchi}}]{airapetian-2016}
{Airapetian}, V.~S., {Glocer}, A., {Gronoff}, G., {H{\'e}brard}, E., \& {Danchi}, W. 2016, Nature Geoscience, 9, 452

\bibitem[{{Arakawa} \& {Kokubo}(2023)}]{arakawa-2023}
{Arakawa}, S. \& {Kokubo}, E. 2023, \aap, 670, A105

\bibitem[{{Banjac} {et~al.}(2019){Banjac}, {Herbst}, \& {Heber}}]{banjac-2019}
{Banjac}, S., {Herbst}, K., \& {Heber}, B. 2019, Journal of Geophysical Research (Space Physics), 124, 50

\bibitem[{{Barth} {et~al.}(2021){Barth}, {Helling}, {St{\"u}eken}, {Bourrier}, {Mayne}, {Rimmer}, {Jardine}, {Vidotto}, {Wheatley}, \& {Fares}}]{barth-2021}
{Barth}, P., {Helling}, C., {St{\"u}eken}, E.~E., {et~al.} 2021, \mnras, 502, 6201

\bibitem[{{Bell}(1978)}]{bell-1978}
{Bell}, A.~R. 1978, \mnras, 182, 147

\bibitem[{{Bellotti} {et~al.}(2023){Bellotti}, {Fares}, {Vidotto}, {Morin}, {Petit}, {Hussain}, {Bourrier}, {Donati}, {Moutou}, \& {H{\'e}brard}}]{bellotti-2023}
{Bellotti}, S., {Fares}, R., {Vidotto}, A.~A., {et~al.} 2023, \aap, 676, A139

\bibitem[{Benner {et~al.}(2020)Benner, Bell, Biondi, Brasser, Carell, Kim, Mojzsis, Omran, Pasek, \& Trail}]{benner-2020}
Benner, S.~A., Bell, E.~A., Biondi, E., {et~al.} 2020, ChemSystemsChem, 2, e1900035

\bibitem[{{Benner} {et~al.}(2019){Benner}, {Kim}, \& {Biondi}}]{benner-2019}
{Benner}, S.~A., {Kim}, H.-J., \& {Biondi}, E. 2019, Life, 9, 84

\bibitem[{Brasser {et~al.}(2016)Brasser, Mojzsis, Werner, Matsumura, \& Ida}]{brasser-2016}
Brasser, R., Mojzsis, S., Werner, S., Matsumura, S., \& Ida, S. 2016, Earth and Planetary Science Letters, 455, 85–93

\bibitem[{{B{\"u}sching} \& {Potgieter}(2008)}]{busching-2008}
{B{\"u}sching}, I. \& {Potgieter}, M.~S. 2008, Advances in Space Research, 42, 504

\bibitem[{{Carolan} {et~al.}(2019){Carolan}, {Vidotto}, {Loesch}, \& {Coogan}}]{carolan-2019}
{Carolan}, S., {Vidotto}, A.~A., {Loesch}, C., \& {Coogan}, P. 2019, \mnras, 489, 5784

\bibitem[{{Claringbold} {et~al.}(2023){Claringbold}, {Rimmer}, {Rugheimer}, \& {Shorttle}}]{claringbold-2023}
{Claringbold}, A.~B., {Rimmer}, P.~B., {Rugheimer}, S., \& {Shorttle}, O. 2023, \aj, 166, 39

\bibitem[{Corti {et~al.}(2016)Corti, Bindi, Consolandi, \& Whitman}]{corti-2016}
Corti, C., Bindi, V., Consolandi, C., \& Whitman, K. 2016, \apj, 829, 8

\bibitem[{{Dyrek} {et~al.}(2024){Dyrek}, {Min}, {Decin}, {Bouwman}, {Crouzet}, {Molli{\`e}re}, {Lagage}, {Konings}, {Tremblin}, {G{\"u}del}, {Pye}, {Waters}, {Henning}, {Vandenbussche}, {Ardevol Martinez}, {Argyriou}, {Ducrot}, {Heinke}, {van Looveren}, {Absil}, {Barrado}, {Baudoz}, {Boccaletti}, {Cossou}, {Coulais}, {Edwards}, {Gastaud}, {Glasse}, {Glauser}, {Greene}, {Kendrew}, {Krause}, {Lahuis}, {Mueller}, {Olofsson}, {Patapis}, {Rouan}, {Royer}, {Scheithauer}, {Waldmann}, {Whiteford}, {Colina}, {van Dishoeck}, {{\"O}stlin}, {Ray}, \& {Wright}}]{dyrek-2024}
{Dyrek}, A., {Min}, M., {Decin}, L., {et~al.} 2024, \nat, 625, 51

\bibitem[{Engelbrecht {et~al.}(2024)Engelbrecht, Herbst, Strauss, Scherer, Light, \& Moloto}]{engelbrecht-2024}
Engelbrecht, N.~E., Herbst, K., Strauss, R. D.~T., {et~al.} 2024, The Astrophysical Journal, 964, 89

\bibitem[{{Engelbrecht} \& {Moloto}(2021)}]{engelbrecht-2021}
{Engelbrecht}, N.~E. \& {Moloto}, K.~D. 2021, \apj, 908, 167

\bibitem[{Finlay {et~al.}(2010)Finlay, Maus, Beggan, Bondar, Chambodut, Chernova, Chulliat, Golovkov, Hamilton, Hamoudi, Holme, Hulot, Kuang, Langlais, Lesur, Lowes, Lühr, Macmillan, Mandea, McLean, Manoj, Menvielle, Michaelis, Olsen, Rauberg, Rother, Sabaka, Tangborn, Tøffner-Clausen, Thébault, Thomson, Wardinski, Wei, \& Zvereva}]{finlay-2010}
Finlay, C.~C., Maus, S., Beggan, C.~D., {et~al.} 2010, Geophysical Journal International, 183, 1216

\bibitem[{{Gallet} \& {Bouvier}(2013)}]{gallet-bouvier-2013}
{Gallet}, F. \& {Bouvier}, J. 2013, \aap, 556, A36

\bibitem[{{Gardner} {et~al.}(2006){Gardner}, {Mather}, {Clampin}, {Doyon}, {Greenhouse}, {Hammel}, {Hutchings}, {Jakobsen}, {Lilly}, {Long}, {Lunine}, {McCaughrean}, {Mountain}, {Nella}, {Rieke}, {Rieke}, {Rix}, {Smith}, {Sonneborn}, {Stiavelli}, {Stockman}, {Windhorst}, \& {Wright}}]{gardner-2006}
{Gardner}, J.~P., {Mather}, J.~C., {Clampin}, M., {et~al.} 2006, \ssr, 123, 485

\bibitem[{Gargaud {et~al.}(2013)Gargaud, Martin, Lopez-Garcia, Montmerle, \& Pascal}]{book-gargaud}
Gargaud, M., Martin, H., Lopez-Garcia, P., Montmerle, T., \& Pascal, R. 2013, Young Sun, Early Earth and the Origins of Life, 299

\bibitem[{{Globus} \& {Blandford}(2020)}]{globus-2020}
{Globus}, N. \& {Blandford}, R.~D. 2020, \apjl, 895, L11

\bibitem[{{G{\"u}nther} {et~al.}(2020){G{\"u}nther}, {Zhan}, {Seager}, {Rimmer}, {Ranjan}, {Stassun}, {Oelkers}, {Daylan}, {Newton}, {Kristiansen}, {Olah}, {Gillen}, {Rappaport}, {Ricker}, {Vanderspek}, {Latham}, {Winn}, {Jenkins}, {Glidden}, {Fausnaugh}, {Levine}, {Dittmann}, {Quinn}, {Krishnamurthy}, \& {Ting}}]{gunther-2020}
{G{\"u}nther}, M.~N., {Zhan}, Z., {Seager}, S., {et~al.} 2020, \aj, 159, 60

\bibitem[{{Helling} \& {Rimmer}(2019)}]{helling-2019}
{Helling}, C. \& {Rimmer}, P.~B. 2019, Philosophical Transactions of the Royal Society of London Series A, 377, 20180398

\bibitem[{{Herbst} {et~al.}(2022){Herbst}, {Baalmann}, {Bykov}, {Engelbrecht}, {Ferreira}, {Izmodenov}, {Korolkov}, {Levenfish}, {Linsky}, {Meyer}, {Scherer}, \& {Strauss}}]{herbst-2022}
{Herbst}, K., {Baalmann}, L.~R., {Bykov}, A., {et~al.} 2022, \ssr, 218, 29

\bibitem[{{Herbst} {et~al.}(2024){Herbst}, {Bartenschlager}, {Grenfell}, {Iro}, {Sinnhuber}, {Taysum}, {Wunderlich}, {Engelbrecht}, {Light}, {Moloto}, {Harre}, {Rauer}, \& {Schreier}}]{herbst-2024}
{Herbst}, K., {Bartenschlager}, A., {Grenfell}, J.~L., {et~al.} 2024, \apj, 961, 164

\bibitem[{{Herbst} {et~al.}(2019){Herbst}, {Grenfell}, {Sinnhuber}, {Rauer}, {Heber}, {Banjac}, {Scheucher}, {Schmidt}, {Gebauer}, {Lehmann}, \& {Schreier}}]{herbst-2019}
{Herbst}, K., {Grenfell}, J.~L., {Sinnhuber}, M., {et~al.} 2019, \aap, 631, A101

\bibitem[{Herbst {et~al.}(2013)Herbst, Kopp, \& Heber}]{herbst-2013}
Herbst, K., Kopp, A., \& Heber, B. 2013, Annales Geophysicae, 31, 1637

\bibitem[{{Herbst} {et~al.}(2017){Herbst}, {Muscheler}, \& {Heber}}]{herbst-2017}
{Herbst}, K., {Muscheler}, R., \& {Heber}, B. 2017, Journal of Geophysical Research (Space Physics), 122, 23

\bibitem[{{Herbst} {et~al.}(2020){Herbst}, {Scherer}, {Ferreira}, {Baalmann}, {Engelbrecht}, {Fichtner}, {Kleimann}, {Strauss}, {Moeketsi}, \& {Mohamed}}]{herbst-2020}
{Herbst}, K., {Scherer}, K., {Ferreira}, S. E.~S., {et~al.} 2020, \apjl, 897, L27

\bibitem[{{Hillas}(1984)}]{hillas-1984}
{Hillas}, A.~M. 1984, \araa, 22, 425

\bibitem[{Itcovitz {et~al.}(2022)Itcovitz, Rae, Citron, Stewart, Sinclair, Rimmer, \& Shorttle}]{itcovitz-2022}
Itcovitz, J.~P., Rae, A. S.~P., Citron, R.~I., {et~al.} 2022, The Planetary Science Journal, 3, 115

\bibitem[{{Ivanova} \& {Taam}(2003)}]{ivanova-2003}
{Ivanova}, N. \& {Taam}, R.~E. 2003, \apj, 599, 516

\bibitem[{{Johnstone} {et~al.}(2021){Johnstone}, {Bartel}, \& {G{\"u}del}}]{johnstone-2021}
{Johnstone}, C.~P., {Bartel}, M., \& {G{\"u}del}, M. 2021, \aap, 649, A96

\bibitem[{{Johnstone} {et~al.}(2015){Johnstone}, {G{\"u}del}, {Brott}, \& {L{\"u}ftinger}}]{johnstone-2015}
{Johnstone}, C.~P., {G{\"u}del}, M., {Brott}, I., \& {L{\"u}ftinger}, T. 2015, \aap, 577, A28

\bibitem[{Kobayashi {et~al.}(2023)Kobayashi, Ise, Aoki, Kinoshita, Naito, Udo, Kunwar, Takahashi, Shibata, Mita, Fukuda, Oguri, Kawamura, Kebukawa, \& Airapetian}]{kobayashi-2023}
Kobayashi, K., Ise, J.-i., Aoki, R., {et~al.} 2023, Life, 13

\bibitem[{{Light} {et~al.}(2025){Light}, {Engelbrecht}, {Herbst}, \& {Scherer}}]{light-2025}
{Light}, J., {Engelbrecht}, N.~E., {Herbst}, K., \& {Scherer}, K.~D. 2025, \mnras, 537, 2097

\bibitem[{{Madhusudhan} {et~al.}(2021){Madhusudhan}, {Piette}, \& {Constantinou}}]{madhusudhan-2021}
{Madhusudhan}, N., {Piette}, A. A.~A., \& {Constantinou}, S. 2021, \apj, 918, 1

\bibitem[{{Madhusudhan} {et~al.}(2023){Madhusudhan}, {Sarkar}, {Constantinou}, {Holmberg}, {Piette}, \& {Moses}}]{madhusudhan-2023}
{Madhusudhan}, N., {Sarkar}, S., {Constantinou}, S., {et~al.} 2023, \apjl, 956, L13

\bibitem[{{McComas} {et~al.}(2008){McComas}, {Ebert}, {Elliott}, {Goldstein}, {Gosling}, {Schwadron}, \& {Skoug}}]{mccomas-2008}
{McComas}, D.~J., {Ebert}, R.~W., {Elliott}, H.~A., {et~al.} 2008, \grl, 35, L18103

\bibitem[{{Mesquita} {et~al.}(2021){Mesquita}, {Rodgers-Lee}, \& {Vidotto}}]{mesquita-2021}
{Mesquita}, A.~L., {Rodgers-Lee}, D., \& {Vidotto}, A.~A. 2021, \mnras, 505, 1817

\bibitem[{{Mesquita} {et~al.}(2022){Mesquita}, {Rodgers-Lee}, {Vidotto}, \& {Kavanagh}}]{mesquita-2022}
{Mesquita}, A.~L., {Rodgers-Lee}, D., {Vidotto}, A.~A., \& {Kavanagh}, R.~D. 2022, \mnras, 515, 1218

\bibitem[{{Miller} \& {Schlesinger}(1983)}]{miller-1983}
{Miller}, S.~L. \& {Schlesinger}, G. 1983, Advances in Space Research, 3, 47

\bibitem[{Miller \& Urey(1959)}]{miller-urey-1959}
Miller, S.~L. \& Urey, H.~C. 1959, Science, 130, 245

\bibitem[{Neher(1971)}]{neher-1971}
Neher, H.~V. 1971, \jgr, 76, 1637

\bibitem[{{{\'O} Fionnag{\'a}in} \& {Vidotto}(2018)}]{dualta-2018}
{{\'O} Fionnag{\'a}in}, D. \& {Vidotto}, A.~A. 2018, \mnras, 476, 2465

\bibitem[{{Padovani} {et~al.}(2009){Padovani}, {Galli}, \& {Glassgold}}]{padovani-2009}
{Padovani}, M., {Galli}, D., \& {Glassgold}, A.~E. 2009, \aap, 501, 619

\bibitem[{{Padovani} {et~al.}(2018){Padovani}, {Ivlev}, {Galli}, \& {Caselli}}]{padovani-2018}
{Padovani}, M., {Ivlev}, A.~V., {Galli}, D., \& {Caselli}, P. 2018, \aap, 614, A111

\bibitem[{{Padovani} {et~al.}(2016){Padovani}, {Marcowith}, {Hennebelle}, \& {Ferri{\`e}re}}]{padovani-2016}
{Padovani}, M., {Marcowith}, A., {Hennebelle}, P., \& {Ferri{\`e}re}, K. 2016, \aap, 590, A8

\bibitem[{{Parker}(1965)}]{parker-1965}
{Parker}, E.~N. 1965, \planss, 13, 9

\bibitem[{Pilchowski {et~al.}(2010)Pilchowski, A, Herbst, \& Heber}]{pilchowski-2010}
Pilchowski, J., A, K., Herbst, K., \& Heber, B. 2010, Astrophysics and Space Sciences Transactions (ASTRA), 6

\bibitem[{{Potgieter}(2013)}]{potgieter-2013}
{Potgieter}, M.~S. 2013, Living Reviews in Solar Physics, 10, 3

\bibitem[{{Reeves} {et~al.}(1992){Reeves}, {Cayton}, {Gary}, \& {Belian}}]{reeves-1992}
{Reeves}, G.~D., {Cayton}, T.~E., {Gary}, S.~P., \& {Belian}, R.~D. 1992, \jgr, 97, 6219

\bibitem[{{Rigby} {et~al.}(2023){Rigby}, {Perrin}, {McElwain}, {Kimble}, {Friedman}, {Lallo}, {Doyon}, {Feinberg}, {Ferruit}, {Glasse}, {Rieke}, {Rieke}, {Wright}, {Willott}, {Colon}, {Milam}, {Neff}, {Stark}, {Valenti}, {Abell}, {Abney}, {Abul-Huda}, {Acton}, {Adams}, {Adler}, {Aguilar}, {Ahmed}, {Albert}, {Alberts}, {Aldridge}, {Allen}, {Altenburg}, {{\'A}lvarez-M{\'a}rquez}, {Alves de Oliveira}, {Andersen}, {Anderson}, {Anderson}, {Argyriou}, {Armstrong}, {Arribas}, {Artigau}, {Arvai}, {Atkinson}, {Bacon}, {Bair}, {Banks}, {Barrientes}, {Barringer}, {Bartosik}, {Bast}, {Baudoz}, {Beatty}, {Bechtold}, {Beck}, {Bergeron}, {Bergkoetter}, {Bhatawdekar}, {Birkmann}, {Blazek}, {Blome}, {Boccaletti}, {B{\"o}ker}, {Boia}, {Bonaventura}, {Bond}, {Bosley}, {Boucarut}, {Bourque}, {Bouwman}, {Bower}, {Bowers}, {Boyer}, {Bradley}, {Brady}, {Braun}, {Breda}, {Bresnahan}, {Bright}, {Britt}, {Bromenschenkel}, {Brooks}, {Brooks}, {Brown}, {Brown}, {Brown}, {Bunker}, {Burger}, {Bushouse}, {Cale}, {Cameron}, {Cameron},
  {Canipe}, {Caplinger}, {Caputo}, {Cara}, {Carey}, {Carniani}, {Carrasquilla}, {Carruthers}, {Case}, {Catherine}, {Chance}, {Chapman}, {Charlot}, {Charlow}, {Chayer}, {Chen}, {Cherinka}, {Chichester}, {Chilton}, {Chonis}, {Clampin}, {Clark}, {Clark}, {Coe}, {Coleman}, {Comber}, {Comeau}, {Connolly}, {Cooper}, {Cooper}, {Coppock}, {Correnti}, {Cossou}, {Coulais}, {Coyle}, {Cracraft}, {Curti}, {Cuturic}, {Davis}, {Davis}, {Dean}, {DeLisa}, {deMeester}, {Dencheva}, {Dencheva}, {DePasquale}, {Deschenes}, {Hunor Detre}, {Diaz}, {Dicken}, {DiFelice}, {Dillman}, {Dixon}, {Doggett}, {Donaldson}, {Douglas}, {DuPrie}, {Dupuis}, {Durning}, {Easmin}, {Eck}, {Edeani}, {Egami}, {Ehrenwinkler}, {Eisenhamer}, {Eisenhower}, {Elie}, {Elliott}, {Elliott}, {Ellis}, {Engesser}, {Espinoza}, {Etienne}, {Etxaluze}, {Falini}, {Feeney}, {Ferry}, {Filippazzo}, {Fincham}, {Fix}, {Flagey}, {Florian}, {Flynn}, {Fontanella}, {Ford}, {Forshay}, {Fox}, {Franz}, {Fu}, {Fullerton}, {Galkin}, {Galyer}, {Garc{\'\i}a Mar{\'\i}n}, {Gardner},
  {Gardner}, {Garland}, {Garrett}, {Gasman}, {Gaspar}, {Gaudreau}, {Gauthier}, {Geers}, {Geithner}, {Gennaro}, {Giardino}, {Girard}, {Giuliano}, {Glassmire}, {Glauser}, {Glazer}, {Godfrey}, {Golimowski}, {Gollnitz}, {Gong}, {Gonzaga}, {Gordon}, {Gordon}, {Goudfrooij}, {Greene}, {Greenhouse}, {Grimaldi}, {Groebner}, {Grundy}, {Guillard}, {Gutman}, {Ha}, {Haderlein}, {Hagedorn}, {Hainline}, {Haley}, {Hami}, {Hamilton}, {Hammel}, {Hansen}, {Harkins}, {Harr}, {Hart}, {Hart}, {Hartig}, {Hashimoto}, {Haskins}, {Hathaway}, {Havey}, {Hayden}, {Hecht}, {Heller-Boyer}, {Henriques}, {Henry}, {Hermann}, {Hernandez}, {Hesman}, {Hicks}, {Hilbert}, {Hines}, {Hoffman}, {Holfeltz}, {Holler}, {Hoppa}, {Hott}, {Howard}, {Howard}, {Hunter}, {Hunter}, {Hurst}, {Husemann}, {Hustak}, {Ilinca Ignat}, {Illingworth}, {Irish}, {Jackson}, {Jahromi}, {Jakobsen}, {James}, {James}, {Januszewski}, {Jenkins}, {Jirdeh}, {Johnson}, {Johnson}, {Jones}, {Jones}, {Jones}, {Jones}, {Jordan}, {Jordan}, {Jurczyk}, {Jurling}, {Kaleida}, {Kalmanson},
  {Kammerer}, {Kang}, {Kao}, {Karakla}, {Kavanagh}, {Kelly}, {Kendrew}, {Kennedy}, {Kenny}, {Keski-kuha}, {Keyes}, {Kidwell}, {Kinzel}, {Kirk}, {Kirkpatrick}, {Kirshenblat}, {Klaassen}, {Knapp}, {Knight}, {Knollenberg}, {Koehler}, {Koekemoer}, {Kovacs}, {Kulp}, {Kumari}, {Kyprianou}, {La Massa}, {Labador}, {Labiano}, {Lagage}, {Lajoie}, {Lallo}, {Lam}, {Lamb}, {Lambros}, {Lampenfield}, {Langston}, {Larson}, {Law}, {Lawrence}, {Lee}, {Leisenring}, {Lepo}, {Leveille}, {Levenson}, {Levine}, {Levy}, {Lewis}, {Lewis}, {Libralato}, {Lightsey}, {Link}, {Liu}, {Lo}, {Lockwood}, {Logue}, {Long}, {Long}, {Loomis}, {Lopez-Caniego}, {Lorenzo Alvarez}, {Love-Pruitt}, {Lucy}, {Luetzgendorf}, {Maghami}, {Maiolino}, {Major}, {Malla}, {Malumuth}, {Manjavacas}, {Mannfolk}, {Marrione}, {Marston}, {Martel}, {Maschmann}, {Masci}, {Masciarelli}, {Maszkiewicz}, {Mather}, {McKenzie}, {McLean}, {McMaster}, {Melbourne}, {Mel{\'e}ndez}, {Menzel}, {Merz}, {Meyett}, {Meza}, {Miskey}, {Misselt}, {Moller}, {Morrison}, {Morse}, {Moseley},
  {Mosier}, {Mountain}, {Mueckay}, {Mueller}, {Mullally}, {Murphy}, {Murray}, {Murray}, {Mustelier}, {Muzerolle}, {Mycroft}, {Myers}, {Myrick}, {Nanavati}, {Nance}, {Nayak}, {Naylor}, {Nelan}, {Nickson}, {Nielson}, {Nieto-Santisteban}, {Nikolov}, {Noriega-Crespo}, {O'Shaughnessy}, {O'Sullivan}, {Ochs}, {Ogle}, {Oleszczuk}, {Olmsted}, {Osborne}, {Ottens}, {Owens}, {Pacifici}, {Pagan}, {Page}, {Park}, {Parrish}, {Patapis}, {Paul}, {Pauly}, {Pavlovsky}, {Pedder}, {Peek}, {Pena-Guerrero}, {Penanen}, {Perez}, {Perna}, {Perriello}, {Phillips}, {Pietraszkiewicz}, {Pinaud}, {Pirzkal}, {Pitman}, {Piwowar}, {Platais}, {Player}, {Plesha}, {Pollizi}, {Polster}, {Pontoppidan}, {Porterfield}, {Proffitt}, {Pueyo}, {Pulliam}, {Quirt}, {Quispe Neira}, {Ramos Alarcon}, {Ramsay}, {Rapp}, {Rapp}, {Rauscher}, {Ravindranath}, {Rawle}, {Regan}, {Reichard}, {Reis}, {Ressler}, {Rest}, {Reynolds}, {Rhue}, {Richon}, {Rickman}, {Ridgaway}, {Ritchie}, {Rix}, {Robberto}, {Robinson}, {Robinson}, {Robinson}, {Rock}, {Rodriguez}, {Rodriguez
  Del Pino}, {Roellig}, {Rohrbach}, {Roman}, {Romelfanger}, {Rose}, {Roteliuk}, {Roth}, {Rothwell}, {Rowlands}, {Roy}, {Royer}, {Royle}, {Rui}, {Rumler}, {Runnels}, {Russ}, {Rustamkulov}, {Ryden}, {Ryer}, {Sabata}, {Sabatke}, {Sabbi}, {Samuelson}, {Sapp}, {Sappington}, {Sargent}, {Sauer}, {Scheithauer}, {Schlawin}, {Schlitz}, {Schmitz}, {Schneider}, {Schreiber}, {Schulze}, {Schwab}, {Scott}, {Sembach}, {Shanahan}, {Shaughnessy}, {Shaw}, {Shawger}, {Shay}, {Sheehan}, {Shen}, {Sherman}, {Shiao}, {Shih}, {Shivaei}, {Sienkiewicz}, {Sing}, {Sirianni}, {Sivaramakrishnan}, {Skipper}, {Sloan}, {Slocum}, {Slowinski}, {Smith}, {Smith}, {Smith}, {Smith}, {Snyder}, {Soh}, {Sohn}, {Soto}, {Spencer}, {Stallcup}, {Stansberry}, {Starr}, {Starr}, {Stewart}, {Stiavelli}, {Straughn}, {Strickland}, {Stys}, {Summers}, {Sun}, {Sunnquist}, {Swade}, {Swam}, {Swaters}, {Swoish}, {Taylor}, {Taylor}, {Te Plate}, {Tea}, {Teague}, {Telfer}, {Temim}, {Thatte}, {Thompson}, {Thompson}, {Thomson}, {Tikkanen}, {Tippet}, {Todd}, {Toolan},
  {Tran}, {Trejo}, {Truong}, {Tsukamoto}, {Tustain}, {Tyra}, {Ubeda}, {Underwood}, {Uzzo}, {Van Campen}, {Vandal}, {Vandenbussche}, {Vila}, {Volk}, {Wahlgren}, {Waldman}, {Walker}, {Wander}, {Warfield}, {Warner}, {Wasiak}, {Watkins}, {Weaver}, {Weilert}, {Weiser}, {Weiss}, {Weissman}, {Welty}, {West}, {Wheate}, {Wheatley}, {Wheeler}, {White}, {Whiteaker}, {Whitehouse}, {Whiteleather}, {Whitman}, {Williams}, {Willmer}, {Willoughby}, {Wilson}, {Wirth}, {Wislowski}, {Wolf}, {Wolfe}, {Wolff}, {Workman}, {Wright}, {Wu}, {Wu}, {Wymer}, {Yates}, {Yeager}, {Yeates}, {Yerger}, {Yoon}, {Young}, {Yu}, {Zak}, {Zeidler}, {Zhou}, {Zielinski}, {Zincke}, \& {Zonak}}]{rigby-2023}
{Rigby}, J., {Perrin}, M., {McElwain}, M., {et~al.} 2023, \pasp, 135, 048001

\bibitem[{Rimmer(2023)}]{rimmer-2023}
Rimmer, P.~B. 2023, Origins of Life on Exoplanets (John Wiley \& Sons, Ltd), 407--424

\bibitem[{{Rimmer} \& {Helling}(2013)}]{rimmer-2013}
{Rimmer}, P.~B. \& {Helling}, C. 2013, \apj, 774, 108

\bibitem[{{Rimmer} \& {Helling}(2016)}]{rimmer-2016}
{Rimmer}, P.~B. \& {Helling}, C. 2016, \apjs, 224, 9

\bibitem[{{Rimmer} {et~al.}(2012){Rimmer}, {Herbst}, {Morata}, \& {Roueff}}]{rimmer-2012}
{Rimmer}, P.~B., {Herbst}, E., {Morata}, O., \& {Roueff}, E. 2012, \aap, 537, A7

\bibitem[{{Rodgers-Lee} {et~al.}(2023){Rodgers-Lee}, {Rimmer}, {Vidotto}, {Louca}, {Taylor}, {Mesquita}, {Miguel}, {Venot}, {Helling}, {Barth}, \& {Lacy}}]{drl-2023}
{Rodgers-Lee}, D., {Rimmer}, P.~B., {Vidotto}, A.~A., {et~al.} 2023, \mnras, 521, 5880

\bibitem[{Rodgers-Lee {et~al.}(2021)Rodgers-Lee, Taylor, Vidotto, \& Downes}]{drl-2021}
Rodgers-Lee, D., Taylor, A.~M., Vidotto, A.~A., \& Downes, T.~P. 2021, \mnras, 504, 1519–1530

\bibitem[{Rodgers-Lee {et~al.}(2020)Rodgers-Lee, Vidotto, Taylor, Rimmer, \& Downes}]{drl-2020}
Rodgers-Lee, D., Vidotto, A.~A., Taylor, A.~M., Rimmer, P.~B., \& Downes, T.~P. 2020, \mnras, 499, 2124–2137

\bibitem[{Rudd {et~al.}(1985)Rudd, Kim, Madison, \& Gallagher}]{rudd-1985}
Rudd, M.~E., Kim, Y.~K., Madison, D.~H., \& Gallagher, J.~W. 1985, Reviews of Modern Physics, 57, 965–994

\bibitem[{Saxena {et~al.}(2019)Saxena, Killen, Airapetian, Petro, Curran, \& Mandell}]{saxena-2019}
Saxena, P., Killen, R.~M., Airapetian, V., {et~al.} 2019, The Astrophysical Journal Letters, 876, L16

\bibitem[{{Scherer} {et~al.}(2025){Scherer}, {Herbst}, {Engelbrecht}, {Ferreira}, {Kleimann}, \& {Light}}]{scherer-2024}
{Scherer}, K., {Herbst}, K., {Engelbrecht}, N.~E., {et~al.} 2025, \aap, 694, A106

\bibitem[{Shea {et~al.}(1987)Shea, Smart, \& Gentile}]{shea-1986}
Shea, M., Smart, D., \& Gentile, L. 1987, Physics of the Earth and Planetary Interiors, 48, 200

\bibitem[{{Sinnhuber} {et~al.}(2012){Sinnhuber}, {Nieder}, \& {Wieters}}]{sinnhuber-2012}
{Sinnhuber}, M., {Nieder}, H., \& {Wieters}, N. 2012, Surveys in Geophysics, 33, 1281

\bibitem[{{Smart} {et~al.}(2000){Smart}, {Shea}, \& {Fl{\"u}ckiger}}]{smart-2000}
{Smart}, D.~F., {Shea}, M.~A., \& {Fl{\"u}ckiger}, E.~O. 2000, \ssr, 93, 305

\bibitem[{St{\o}rmer(1955)}]{stormer-1955}
St{\o}rmer, C. 1955, The Polar Aurora, International monographs on radio (Clarendon Press)

\bibitem[{Svensmark(2006)}]{svensmark-2006}
Svensmark, H. 2006, Astronomische Nachrichten, 327, 871–875

\bibitem[{Tarduno {et~al.}(2020)Tarduno, Cottrell, Bono, Oda, Davis, Fayek, van~’t Erve, Nimmo, Huang, Thern, Fearn, Mitra, Smirnov, \& Blackman}]{tarduno-2020}
Tarduno, J.~A., Cottrell, R.~D., Bono, R.~K., {et~al.} 2020, Proceedings of the National Academy of Sciences, 117, 2309

\bibitem[{{Tian} {et~al.}(2011){Tian}, {Kasting}, \& {Zahnle}}]{tian-2011}
{Tian}, F., {Kasting}, J.~F., \& {Zahnle}, K. 2011, Earth and Planetary Science Letters, 308, 417

\bibitem[{{T{\'o}th}(1996)}]{toth_1996}
{T{\'o}th}, G. 1996, Astrophysical Letters and Communications, 34, 245

\bibitem[{{Varela} {et~al.}(2023){Varela}, {Brun}, {Strugarek}, {R{\'e}ville}, {Zarka}, \& {Pantellini}}]{varela-2023}
{Varela}, J., {Brun}, A.~S., {Strugarek}, A., {et~al.} 2023, \mnras, 525, 4008

\bibitem[{{Vidotto} {et~al.}(2014){Vidotto}, {Gregory}, {Jardine}, {Donati}, {Petit}, {Morin}, {Folsom}, {Bouvier}, {Cameron}, {Hussain}, {Marsden}, {Waite}, {Fares}, {Jeffers}, \& {do Nascimento}}]{vidotto-2014}
{Vidotto}, A.~A., {Gregory}, S.~G., {Jardine}, M., {et~al.} 2014, \mnras, 441, 2361

\bibitem[{Vos \& Potgieter(2015)}]{vos-2015}
Vos, E.~E. \& Potgieter, M.~S. 2015, \apj, 815, 119

\bibitem[{{Welbanks} \& {Madhusudhan}(2019)}]{madhusudhan-2019}
{Welbanks}, L. \& {Madhusudhan}, N. 2019, \aj, 157, 206

\bibitem[{{Wogan} {et~al.}(2024){Wogan}, {Batalha}, {Zahnle}, {Krissansen-Totton}, {Tsai}, \& {Hu}}]{wogan-2024}
{Wogan}, N.~F., {Batalha}, N.~E., {Zahnle}, K.~J., {et~al.} 2024, \apjl, 963, L7

\bibitem[{{Zahnle} {et~al.}(2020){Zahnle}, {Lupu}, {Catling}, \& {Wogan}}]{zahnle-2020}
{Zahnle}, K.~J., {Lupu}, R., {Catling}, D.~C., \& {Wogan}, N. 2020, The Planetary Science Journal, 1, 11

\end{thebibliography}

\begin{appendix}
    \srr{\section{Pressure-altitude plot}
    \label{subsec:pz-plots}
    \begin{figure}
        \centering
        \includegraphics[width=\hsize]{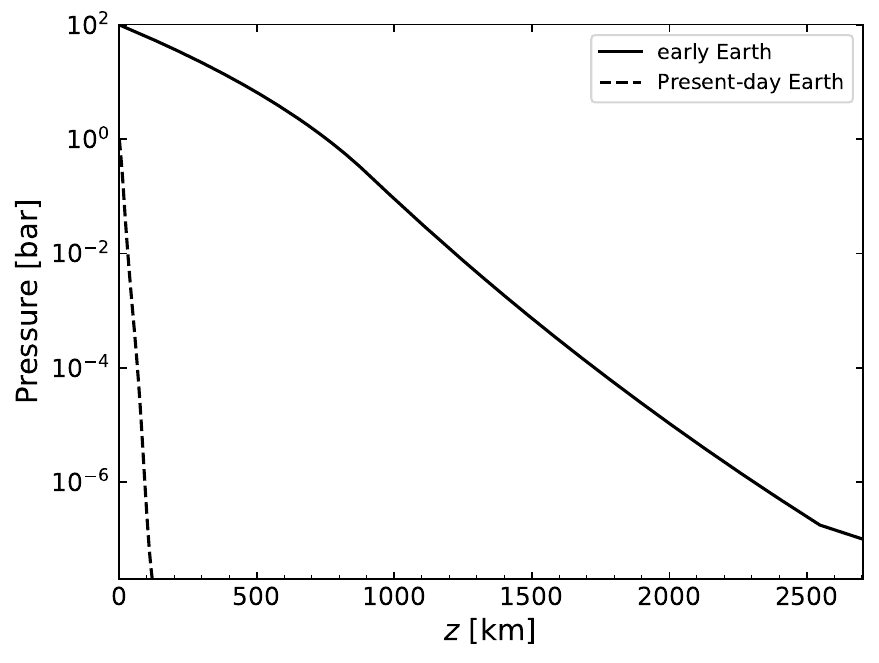}
        \caption{Pressure-altitude plot for a post-impact early Earth atmosphere (solid line) and for the present-day Earth (dashed line).}
        \label{fig:pz}
    \end{figure}
    Fig. \ref{fig:pz} shows pressure as a function of altitude for the post-impact early Earth atmosphere (solid line) and for the present-day Earth atmosphere (dashed line). The steeper decrease in $P$ with increasing $z$ for the present-day Earth atmosphere compared to the early Earth atmosphere is related to the scale height in each case as described in Section \ref{subsec:density-profile}.
    }

    \section{Impact of planetary magnetic field on cosmic ray spectra}
    \label{subsec:bfield}
    Because cosmic rays are charged particles, they are deflected while moving in magnetic fields. There have been numerous studies on the trajectory of cosmic rays in planetary magnetic fields, since Størmer's work on trajectories of charged particles which are either reflected or penetrate Earth's atmosphere \citep{stormer-1955}. Computing the trajectories of particles with a given rigidity, $R = p/q$ (where $p$ is the particle's momentum and $q$ its charge), through a model of the Earth's magnetic field is used to explain the number of reflected or penetrating cosmic rays \citep{shea-1986}. Above a cutoff rigidity, $R_{\rm{C}}$, at a given location, there are no more reflected trajectories. Below $R_{\rm{C}}$, $j(T)$ becomes more suppressed with decreasing $R$ as the number of reflected trajectories increases. To approximate $R_{\rm{C}}$ at a given geomagnetic latitude, $\lambda$, we use the Størmer cutoff rigidity \citep{stormer-1955}, given by:
    \begin{equation}
      R_{\rm{C}} = \frac{M \cos^4\lambda}{4r_{\rm{\oplus}}^2}, \,
      \label{eq:stormer}
   \end{equation}
   where $M$ ($\rm{G \cdot cm^3}$) is the Earth's magnetic dipole moment, $r_{\rm{\oplus}}$ (cm) is the Earth's radius and $R_{\rm{C}}$ is in statvolts. \srr{ This approximation is only valid for $\lambda > 60 ^{\circ}$ \citep{pilchowski-2010}. We calculate $R_{\rm{C}}$ at $\lambda = 60^\circ$ as the extreme limit for this approximation to illustrate the greatest level of suppression.} At higher latitudes as $\lambda \xrightarrow{} 90^{\circ}$, $R_{\rm{C}} \xrightarrow{} 0$ and the unsuppressed spectrum is recovered.\\
   More complex models using trajectory tracing \citep[e.g.][]{smart-2000,herbst-2013} have been used to investigate the effect of the planetary magnetic field on cosmic ray propagation in the present-day Earth atmosphere, including at low latitudes. These studies use models of Earth's magnetic field to calculate $R_{\rm{C}}$ throughout the atmosphere. As a first step, for Earth at 200 Myr, we use the Eq. \ref{eq:stormer} to give an indication of the possible effect of a planetary magnetic field on the cosmic ray spectra in the early Earth atmosphere. \srr{ We use a value of the Earth's surface magnetic field strength in the equatorial region of $B_{\rm{\oplus}} = 0.13 \rm{G}$ for the early Earth\footnote{Here the early Earth refers to Earth during its first $\sim600$ Myr, rather than specifically a post-impact early Earth.} \citep{tarduno-2020}. For comparison, at present in the equatorial region $B_{\rm{\oplus}} \sim 0.3 \rm{G}$ \citep{finlay-2010,varela-2023}. From Eq. \ref{eq:stormer} and $M = B_{\rm{\oplus}}r_{\rm{\oplus}}^3$, we obtain $R_{\rm{C}} = 0.74 \rm{GV}$ for $\lambda = 60^{\circ}$. For protons this corresponds to a cutoff energy, $T_{\rm{C}} = 0.74 \rm{GeV}$.}

    \begin{figure}
   \centering
   \includegraphics[width=\hsize]{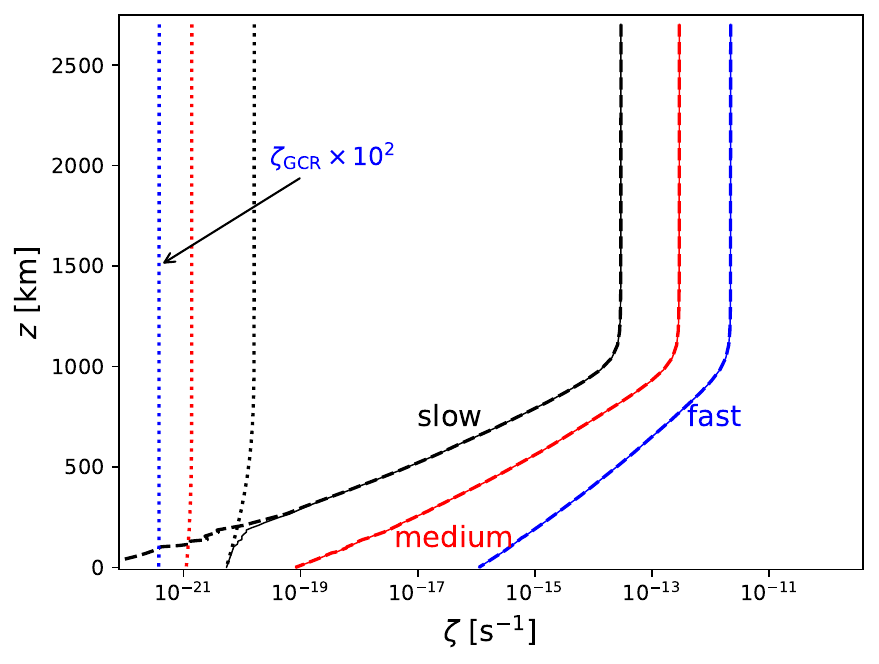}
      \caption{Ionisation rate, $\zeta$, as a function of height in the early Earth atmosphere at 200 Myr due to Galactic and \srr{solar} cosmic  rays, with \srr{$T_{\rm{C}} = 0.74 \rm{GeV}$}. The linestyles are the same as in Fig. \ref{fig:ZetaPlot}. We use the same axes as Fig. \ref{fig:ZetaPlot} for the purpose of comparison.
              }
         \label{fig:ZetaB13}
   \end{figure}

   For $T<T_{\rm{C}}$, $j(T)$ is suppressed but not zero. The suppression of the cosmic ray spectra for $T < T_{\rm{C}}$ depends on both $T$ and $\lambda$. This can be seen in the AMS-01 measurements of cosmic rays taken at different latitudes for $z \sim 320 - 390 \rm{km}$, presented in Fig. 4.9 from \citet{aguilar-2002}. To approximate the suppression of the cosmic ray spectra for $T < T_{\rm{C}}$ shown in \citet{aguilar-2002}, we use the suppression function:
   \begin{equation}
      j_{\rm{supp}}(T) = \frac{j(T)}{2}\left(1-\tanh\left(a\frac{T_{\rm{C}}}{T}-b\right)\right), \,
      \label{eq:supp}
   \end{equation}
   where $a = 3.7 - 2.5 \Theta$, $b = 2$ and $\Theta$ (radians) is the geomagnetic latitude. We then model the propagation of the cosmic rays through the post-impact early Earth atmosphere and calculate $\zeta(z)$ as previously described in Section \ref{subsec:cr-model}. 

   \srr{ Figure \ref{fig:ZetaB13} shows $\zeta(z)$ in the post-impact early Earth atmosphere at 200 Myr for solar and Galactic cosmic rays suppressed by a planetary magnetic field with $B_{\rm{\oplus}} = 0.13 \rm{G}$ and $T_{\rm{C}} = 0.74 \rm{GeV}$. The linestyles are the same as in Fig. \ref{fig:ZetaPlot}. 
   
   For solar cosmic rays in the upper atmosphere the high $\zeta_{\rm{SCR}}$ shown in Fig. \ref{fig:ZetaPlot} due to the high $j(T,z)$ at low energies is decreased by several orders of magnitude in the presence of the planetary magnetic field. For $\Omega = 15 \Omega_{\rm{\odot}}$ at the top of the atmosphere, $\zeta_{\rm{SCR}} = 2\times 10^{-10}\rm{s^{-1}}$, decreasing to $\zeta_{\rm{SCR}} = 2\times 10^{-12}\rm{s^{-1}}$ when the spectrum is suppressed by the planetary magnetic field. Additionally, $\zeta_{\rm{SCR}}$ remains constant deeper into the atmosphere (until $z < 1000 \rm{km}$). For comparison, $\zeta_{\rm{SCR}}$ shown in Fig. \ref{fig:ZetaPlot} begins to decrease with decreasing $z$ when $z < 1700$ km.
   Closer to the surface ($z < 1000 \rm{km}$), $\zeta_{\rm{SCR}}$ is determined by $j(T,z)$ of the higher-energy ($T > T_{\rm{C}}$) cosmic rays. At these higher energies $T > T_{\rm{C}}$ and $\zeta_{\rm{SCR}}$ is unchanged compared to the results presented in Section \ref{subsec:ionisation}. }
    
   \srr{To account for the possibility of a stronger planetary magnetic field for the early Earth we use a value of the Earth's surface magnetic field strength in the equatorial region of $B_{\rm{\oplus}} = 0.45 \rm{G}$ \citep{varela-2023}. For $\lambda = 60^{\circ}$ for protons this corresponds to a cutoff energy, $T_{\rm{C}} = 2.56 \rm{GeV}$. 
   
   }
 
    \begin{figure}
   \centering
   \includegraphics[width=\hsize]{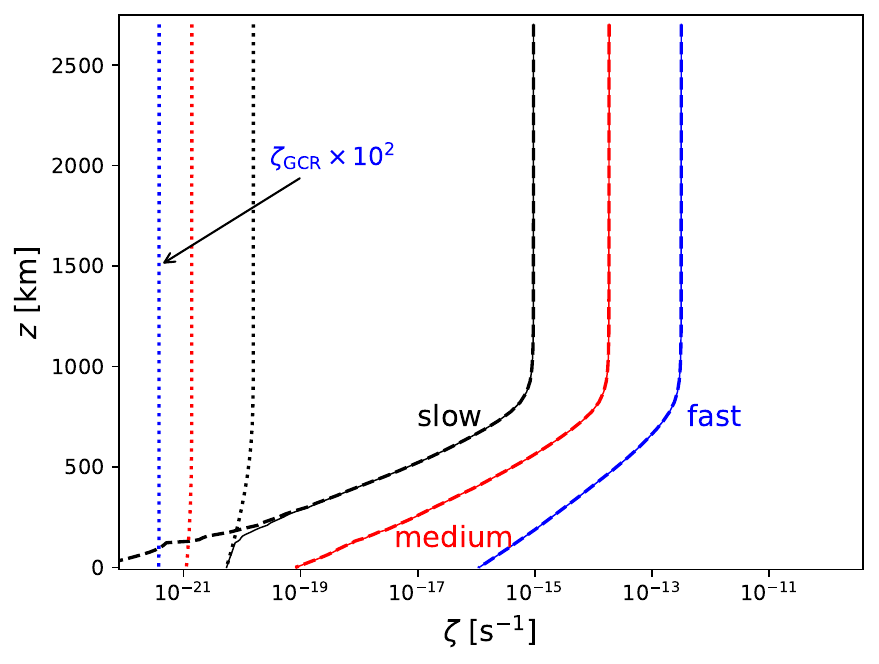}
      \caption{Ionisation rate, $\zeta$, as a function of height in the early Earth atmosphere at 200 Myr due to Galactic and \srr{solar} cosmic  rays, with $T_{\rm{C}} = 2.56 \rm{GeV}$. The linestyles are the same as in Fig. \ref{fig:ZetaPlot}. We use the same axes as Fig. \ref{fig:ZetaPlot} for the purpose of comparison.
              }
         \label{fig:ZetaB}
   \end{figure}
   
   Figure \ref{fig:ZetaB} shows $\zeta(z)$ in the post-impact early Earth atmosphere at 200 Myr for solar and Galactic cosmic rays suppressed by a planetary magnetic field with $B_{\rm{\oplus}} = 0.45 \rm{G}$ and $T_{\rm{C}} = 2.56 \rm{GeV}$. The linestyles are the same as in Fig. \ref{fig:ZetaPlot}. 
   
   For \srr{solar} cosmic rays in the upper atmosphere the high $\zeta_{\rm{SCR}}$ shown in Fig. \ref{fig:ZetaPlot} due to the high $j(T,z)$ at low energies is decreased by several orders of magnitude in the presence of the planetary magnetic field. For \srr{$\Omega = 15 \Omega_{\rm{\odot}}$} at the top of the atmosphere, $\zeta_{\rm{SCR}} = 2\times 10^{-10}\rm{s^{-1}}$, decreasing to $\zeta_{\rm{SCR}} = 1\times 10^{-12}\rm{s^{-1}}$ when the spectrum is suppressed by the planetary magnetic field. Additionally, $\zeta_{\rm{SCR}}$ remains constant deeper into the atmosphere (until $z < 1000 \rm{km}$). For comparison, $\zeta_{\rm{SCR}}$ shown in Fig. \ref{fig:ZetaPlot} begins to decrease with decreasing $z$ when $z < 1700$ km.
   Closer to the surface ($z < 1000 \rm{km}$), $\zeta_{\rm{SCR}}$ is determined by $j(T,z)$ of the higher-energy ($T > T_{\rm{C}}$) cosmic rays. At these higher energies $T > T_{\rm{C}}$ and $\zeta_{\rm{SCR}}$ is unchanged compared to the results presented in Section \ref{subsec:ionisation}. 
   \srr{ Similar to the $T_{\rm{C}} = 0.74 \rm{GeV}$ scenario, $\zeta_{\rm{SCR}}$ at the top of the atmosphere is decreased by several orders of magnitude for $T_{\rm{C}} = 2.56 \rm{GeV}$. For $\Omega = 15 \Omega_{\rm{\odot}}$ at the top of the atmosphere, $\zeta_{\rm{SCR}} = 2\times 10^{-10}\rm{s^{-1}}$, decreasing to $\zeta_{\rm{SCR}} = 3\times 10^{-13}\rm{s^{-1}}$ when $T_{\rm{C}} = 2.56 \rm{GeV}$. At the surface $\zeta_{\rm{SCR}}$ is unchanged compared to the results presented in Section \ref{subsec:ionisation}.
   
   For Galactic cosmic rays for both $T_{\rm{C}} = 0.74 \rm{GeV}$ and $T_{\rm{C}} = 2.56 \rm{GeV}$}, compared to the results presented in Section \ref{subsec:ionisation}, $\zeta_{\rm{GCR}}$ is unchanged by the planetary magnetic field. The Galactic cosmic ray spectra have maximum $j(T,z)$ at high energies ($T > T_{\rm{C}}$) and low $j(T,z)$ for $T \lesssim T_{\rm{C}}$.

   Overall, we find that the suppression of the top-of-atmosphere cosmic ray spectra at 200 Myr by $B_{\rm{\oplus}}$ does not result in a decreased $\zeta$ near the surface compared to the results described in Section \ref{subsec:ionisation}. \\
\end{appendix}

\end{document}